\newcommand{\nii}{[N{\sc{ii}}]}
\newcommand{\lognii}{$\log$([N{\sc{ii}}]/H$\alpha$)}

\newcommand{\cii}{[C{\sc{ii}}]}
\newcommand{\oiii}{[O{\sc{iii}}]}
\newcommand{\ha}{H$\alpha$}
\newcommand{\hb}{H$\beta$}

\newcommand{\sii}{[S{\sc{ii}}]}

\newcommand{\kms}{${\rm km~s^{-1}}$}    
\newcommand{\aless}{ALESS073.1}

\documentclass[longauth]{aa}  

\usepackage{graphicx}
\usepackage{txfonts}
\usepackage{xcolor}
\usepackage{lscape}
\usepackage{xcolor}

\usepackage{txfonts}
\usepackage[colorlinks=true,allcolors=blue]{hyperref}

\begin{document}

   \title{GA-NIFS: Early-stage feedback in a heavily obscured AGN at $z=4.76$}

   \subtitle{}
   \authorrunning{E. Parlanti et al.}

   \author{Eleonora Parlanti \thanks{E-mail: \href{mailto:eleonora.parlanti@sns.it}{eleonora.parlanti@sns.it}}
  \inst{\ref{inst:SNS}} 
  \and  
   Stefano Carniani  \inst{\ref{inst:SNS}}
\and 
Hannah \"Ubler \inst{\ref{inst:kavli}, \ref{inst:cavendish}}
\and
Giacomo Venturi \inst{\ref{inst:SNS}}
\and
Chiara Circosta \inst{\ref{inst:esaspain}, \ref{inst:london}}
\and
Francesco D'Eugenio  \inst{\ref{inst:kavli}, \ref{inst:cavendish}}
\and
Santiago Arribas \inst{\ref{inst:centroastrob}}
\and
Andrew J. Bunker \inst{\ref{inst:oxford}}
\and
Stéphane Charlot\inst{\ref{inst:paris}}
\and 
Nora L\"utzgendorf\inst{\ref{inst:esa}}
\and
Roberto Maiolino \inst{\ref{inst:kavli}, \ref{inst:cavendish}, \ref{inst:london}}
\and
Michele Perna \inst{\ref{inst:centroastrob}}
\and
Bruno Rodr\'iguez Del Pino\inst{\ref{inst:centroastrob}}
\and 
Chris J. Willott \inst{\ref{inst:nrc}}
\and
Torsten Böker \inst{\ref{inst:esa}}
\and
Alex J. Cameron \inst{\ref{inst:oxford}}
\and
Jacopo Chevallard\inst{\ref{inst:oxford}}
\and 
Giovanni Cresci \inst{\ref{inst:arcetri}}
\and 
Gareth C. Jones \inst{\ref{inst:oxford}}
\and
Nimisha Kumari \inst{\ref{inst:aura}}
\and 
Isabella Lamperti\inst{\ref{inst:centroastrob}}
\and
Jan Scholtz \inst{\ref{inst:kavli}, \ref{inst:cavendish}}
   }
    \institute{Scuola Normale Superiore, Piazza dei Cavalieri 7, I-56126 Pisa, Italy \label{inst:SNS} \and 
    Kavli Institute for Cosmology, University of Cambridge, Madingley Road, Cambridge, CB3 0HA, UK \label{inst:kavli} \and
    Cavendish Laboratory - Astrophysics Group, University of Cambridge, 19 JJ Thomson Avenue, Cambridge, CB3 0HE, UK \label{inst:cavendish} 
        \and
        European Space Agency (ESA), European Space Astronomy Centre (ESAC), Camino Bajo del Castillo s/n, 28692 Villanueva de la Cañada, Madrid, Spain \label{inst:esaspain}
    \and
    Department of Physics and Astronomy, University College London, Gower Street, London WC1E 6BT, UK \label{inst:london}
    \and
    Centro de Astrobiolog\'{\i}a (CAB), CSIC-INTA, Ctra. de Ajalvir km 4, Torrej\'on de Ardoz, E-28850, Madrid, Spain \label{inst:centroastrob} 
        \and
    University of Oxford, Department of Physics, Denys Wilkinson Building, Keble Road, Oxford OX13RH, United Kingdom \label{inst:oxford}
    \and
    Sorbonne Universit\'e, CNRS, UMR 7095, Institut d'Astrophysique de Paris, 98 bis bd Arago, 75014 Paris, France \label{inst:paris}
    \and
    European Space Agency, c/o STScI, 3700 San Martin Drive, Baltimore, MD 21218, USA \label{inst:esa}
\and
NRC Herzberg, 5071 West Saanich Rd, Victoria, BC V9E 2E7, Canada \label{inst:nrc}
\and
INAF - Osservatorio Astrofisco di Arcetri, largo E. Fermi 5, 50127 Firenze, Italy\label{inst:arcetri}
\and
AURA for European Space Agency, Space Telescope Science Institute, 3700 San Martin Drive. Baltimore, MD, 21210 \label{inst:aura}
    }
    
   \date{}

 
  \abstract{
Dust-obscured galaxies are thought to represent an early evolutionary phase of massive galaxies in which the active galactic nucleus (AGN) is still deeply buried in significant amounts of dusty material and its emission is strongly suppressed. The unprecedented sensitivity of the \textit{James Webb Space Telescope} (JWST) enabled us for the first time to detect the rest-frame optical emission of heavily obscured AGN and unveil the properties of the hidden accreting super-massive black holes (BHs).  In this work, we present the JWST/NIRSpec integral field spectroscopy (IFS) data of \aless, a massive ($\rm \log(M_{\star}/M_\odot)=10.98$) dusty, star-forming galaxy at $z = 4.755$ hosting an AGN at its center. The detection of a very broad ($>9000$~\kms) \ha\ emission associated with the broad line region (BLR) confirms the presence of a BH ($\log(M_{\rm BH}/\rm M_\odot)>8.7$) accreting at less than 18\% of its Eddington limit. The identification of the BLR classifies the target as a type 1 AGN despite the observed high column density of $N_{\rm H}\sim10^{24}~{\rm cm^{-2}}$.  The rest-frame optical emission lines also reveal a fast ($\sim 1700$~\kms) ionized gas outflow marginally resolved in the galaxy center. The high sensitivity of NIRSpec allowed us to perform the kinematic analysis of the narrow \ha\ component, which indicates that the warm ionized gas velocity field is consistent with disk rotation. Interestingly, we find that in the innermost nuclear regions ($< 1.5$~kpc), the intrinsic velocity dispersion of the disk reaches $\sim150$ \kms, which is $\sim2-3$ times higher than the velocity dispersion inferred from the \cii 158$\mu$m line tracing mostly cold gas. Since at large radii the velocity dispersion of the warm and cold gas are comparable, we conclude that the outflows are injecting turbulence in the warm ionized gas in the central region, but they are not sufficiently powerful to disrupt the dense gas and quench star formation. These findings support the scenario that dust-obscured galaxies represent the evolutionary stage preceding the unobscured quasar when all gas and dust are removed from the host.
  }

   \keywords{quasars: supermassive black holes -- quasars: emission lines -- ISM: jets and outflows -- Galaxies: high-redshift -- Galaxies: kinematics and dynamics.
               }

   \maketitle
%

\section{Introduction}

Supermassive black holes (BHs) are thought to reside in the center of the majority of local massive galaxies \citep{Hopkins:2008}. A symbiotic connection between the growth of BHs and their hosts is suggested by the observed relations between the BH masses and the galaxy properties \citep[e.g.,][]{Kormendy:2013, Heckman:2014}.
Specifically, observations of local galaxies reveal tight relations between the BH mass and the stellar velocity dispersion and the mass and luminosity of the galactic bulge.  These relations hold up throughout several orders of magnitude in black hole masses and galaxy properties and up to high redshift with different normalizations \citep{Carraro:2020, Suh:2020}. 

During the accretion phase, BHs are revealed as active galactic nuclei (AGNs) due to the radiation emitted from radio to X-ray wavelengths by the accretion disk \citep{Padovani:2017}. X-ray surveys show that the most luminous and massive AGNs were most numerous at $z>1$ and, in particular, observations find the cosmic black hole accretion rate density peaks at $z \sim 2$, such as the cosmic star formation rate (SFR) density \citep{Shankar:2009, Delvecchio:2014, Madau:2014, Aird:2015, Ananna:2019, Brandt:2022}. Both the cosmic star formation rate and BH accretion rate are driven by the availability of cold gas in the system \citep{Hopkins:2008}.
The tight relations between the BH and host galaxy properties and the similar evolution of activity with redshift suggests that BHs and the galaxies they inhabit have undergone a common evolutionary process  \citep{Kormendy:2013}. 

The origin and the mechanisms that regulate the coevolution of BHs with their host galaxy are still unclear.  
Massive galactic outflows driven by the radiation emitted by the most luminous AGNs are considered a fundamental physical process in the evolution of galaxies. They are believed to regulate star formation \citep[SF, e.g.,][]{Fabian:2012, Zubovas:2014, Muratov:2015, Yuan:2018, Yoon:2018, Nelson:2019} and reduce the number of galaxies at the high-mass end of the stellar mass function \citep{Benson:2003, Puchwein:2013}. Such fast outflows can potentially accelerate a substantial mass of gas beyond the escape velocity of the local gravitational potential, inject turbulence in the interstellar medium (ISM), and/or heat the gas in the galaxy, in this way damping or even halting SF in their host galaxies (``negative feedback'' ; e.g., \citealt{Fabian:2012, Harrison:2017}).
On the other hand, AGN feedback has been observed to enhance star formation (``positive feedback''; e.g., \citealt{Shin:2019}), with stars actively forming in outflowing material \citep[e.g.,][]{Maiolino:2017, Gallagher:2019}.
It is thus fundamental to investigate the feedback mechanism over the various phases of the evolution of the BHs and galaxies to understand their role in the coevolution process.

Submillimeter galaxies (SMGs) are a class of high-redshift ($z>0.1$) galaxies mostly characterized by a high luminosity in the far infrared (FIR) continuum emission  ($L_{\rm FIR}>10^{11}$~L$_\odot$), high SFRs (SFR $\sim 10^3$ M$_\odot$ yr$^{-1}$), and high dust content ($M_{\rm dust} = 10^{8-10}$; \citealt{Santini:2010}). Although SMGs are a rare cosmological population of galaxies, they account for $\sim$ 20\% of the SFR cosmic density at $1<z<5$ \citep{Swinbank:2014}. These galaxies are also thought to be the precursors of local massive and quiescent early-type galaxies (ETGs) that host BHs with masses $M_{\rm BH}>10^8~{\rm M}_\odot$ in their centers \citep{Sanders:1998,Swinbank:2004, Hopkins:2008, Toft:2014}. Quiescent galaxies are already common at ($z\sim 2\text{--}3$), and are observed up to $z \sim 4.6$ \citep{Carnall:2023}.
The evolutionary path that connects SMGs and quiescent galaxies is thought to be driven by the interaction between the galaxy and the BH \citep{Sanders:1998, Hopkins:2008}. 
The current theoretical scenarios expect that the feeding of galaxies with gas from the cosmic web or via mergers triggers both episodes of intense SF and BH accretion. This process is thought to be self-regulated. 
In particular, the feedback from the accreting BHs should be able to balance the SF in the host galaxy and the accretion of gas on the BH itself due to powerful radiation-driven outflows that are able to sweep out the host galaxy gas reservoir \citep{Debuhr:2012} and halt the accretion of gas \citep{Peng:2015}. These outflows are thought to become important when the accretion rate on the BH reaches the Eddington rate \citep{King:2015}. The dust in the system is also swept out, allowing the radiation coming from the BH accretion zone to be detected and identified as a luminous quasar (QSO) \citep{Sanders:1998, Hopkins:2008}.
The removal and heating of gas due to the AGN feedback halts SF in the system, turning the host into a ``red and dead'' galaxy. This is consistent with what is observed in local ETGs, which are dominated by old stellar populations, with formation redshifts of $z>2$ 
 \citep{Thomas:2005, McDermid:2015}. Several studies also find galaxies whose gas content has been depleted on short timescales \citep{Bezanson:2019, Williams:2021}. Moreover,  recent observations of quiescent galaxies at $z>3$ indicate that fast quenching processes are already in place in the first 2 Gyr of the Universe 
\citep{Glazebrook:2017, Valentino:2020, Carnall:2023}. In conclusion, it is crucial to study the SMG population to asses whether they really represent the evolutionary stage preceding the active QSO phase and, thus, whether they are the progenitors of local ETGs or not. Moreover, SMGs enabled us to study the first phases of the BH feedback process outlined above.

The ISM properties of the high-redshift SMG population have been mainly studied through millimeter observations, which do not allow us to determine whether a BH is hidden at the center of the galaxies. 
Up until the advent of the James Webb Space Telescope (JWST), the rest-frame optical emission lines from galaxies at $z>4$ were extremely difficult to access from ground-based telescopes because the lines are redshifted to wavelength which are outside the atmospheric windows. Thanks to its unique sensitivity, the  NIRSpec instrument \citep{Jakobsen:2022} on board JWST has already proven its capabilities to detect faint $z>4$ galaxies and identify serendipitous AGNs  by observing the emission that arises from the broad line regions (BLRs) surrounding super-massive BHs \citep{Kocevski:2023, Maiolino:2023, Matthee:2023, Ubler:2023}

By using the Integral Field Spectrograph (IFS) mode of  NIRSpec    \citep{Boker:2022} we can also exploit the rest-frame optical lines at high redshift to spatially resolve the emission from the ISM and stellar population and determine the impact of the feedback mechanism on the host galaxy  \citep{Cresci:2023, Marshall:2023, Perna:2023, Ubler:2023}. 

 In this work, we study the properties of \aless\ (from the LABOCA ECDFS submillimeter survey LESS J033229.3-275619), a massive ($M_\star \sim 10^{11}$ M$_{\odot}$) SMG at  $z = 4.755$ showing  X-ray emission and the presence of a highly ionized iron emission line which is consistent with a Compton thick AGN ($N_{\rm H} = 17.0^{+11.7}_{-6.8} \times 10^{23}$~cm$^{-2}$) \citep{Vanzella:2009, Gilli:2011, Circosta:2019, Giallongo:2019}. \aless~ has a plethora of observations targeting both rest-frame far-infrared \citep{Coppin:2009, Debreuck:2011, Nagao:2012, Debreuck:2014, Damato:2020, Zhao:2020,  Lelli:2021}, UV \citep{Vanzella:2006, Vanzella:2009, Coppin:2009}, and X-ray emission \citep{Tozzi:2006, Gilli:2011, Vito:2013, Gilli:2014, Circosta:2019, Giallongo:2019}.
The bright rest-frame far-infrared emission indicates a massive burst of SF with SFR of $\sim 1000$ M$_\odot$ yr$^{-1}$ \citep{Coppin:2009, Gilli:2014, Swinbank:2014, Circosta:2019} and a large dust content ($M_{\rm dust} \sim 10^9$ M$_\odot$; \citealt{Swinbank:2014, daCunha:2015}).
This massive, dusty, highly star-forming SMG offers us a great opportunity to study the interplay between AGN and host galaxy at their first stages when the AGN accretion is creating a weak outflow that is starting to reveal the central AGN. We use NIRSpec observations to study the properties of the BH and investigate the impact of AGN-driven outflows (if any) on the host galaxy. 

This work is structured as follows. In section \ref{sec:obs}, we describe the target and the new JWST IFS observations. In section \ref{sec:dataanalysis}, we present the analysis of the spatially integrated and single-spaxel spectra. In section \ref{sec:bpt}, we investigate which is the primary excitation mechanism of the gas. 
In Section \ref{sec:blackhole} we compute the black hole mass and compare the position of \aless on the $M_\star-M_{\rm BH}$ plane with other low and high redshift AGNs and QSOs.
In section \ref{sec:Outflow}, we study the  properties and  energetics of the outflow. In section \ref{sec:kinematics}, we perform a detailed kinematic analysis of the host galaxy.
We discuss our results in section \ref{sec:discussion}, and we draw our conclusions in section \ref{sec:conclusion}.
In this work, we adopt the cosmological parameters from \cite{Planck:2015}: $H_0$ =  67.7 \kms Mpc$^{-1}$, $\Omega_m$ = 0.307, and $\Omega_\Lambda$ = 0.691, $0.1$\arcsec$=0.66$ kpc at $z=$4.755.

\section{Observations}
\label{sec:obs}

\subsection{Target}

\begin{figure}[ht!]
   \resizebox{\hsize}{!}
    {\includegraphics[width = \hsize]{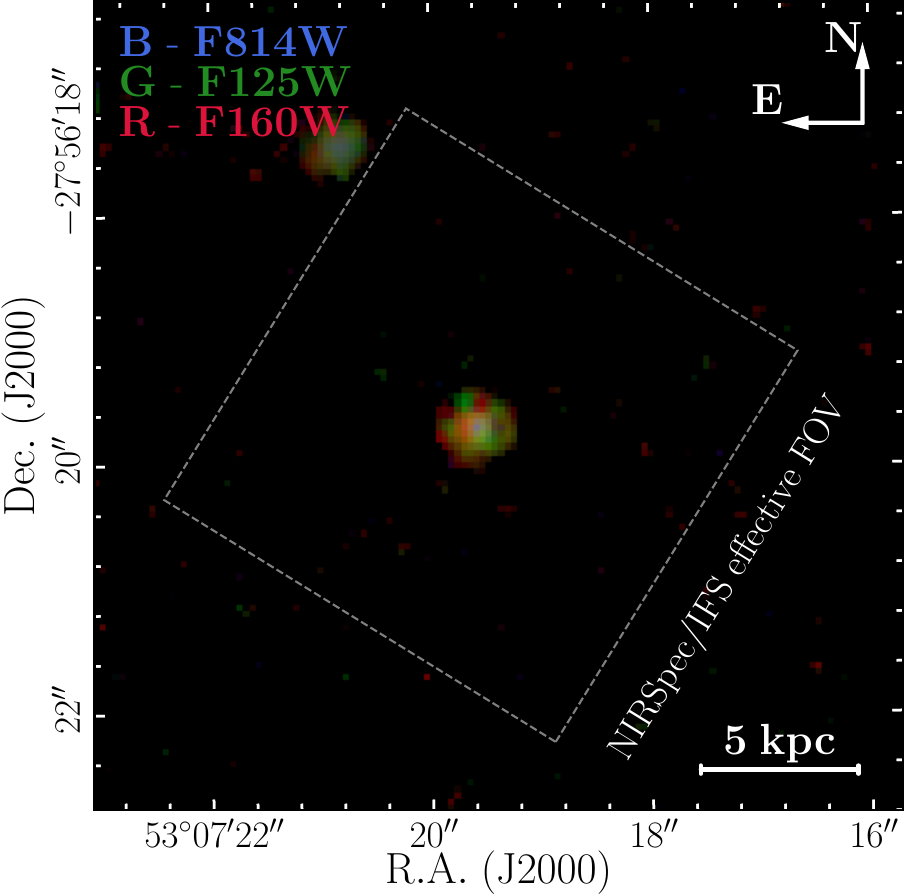}}
    \caption{RGB image of ALESS073.1. The
colors are a combination of HST/ACS and HST/WFC3 images. White contours highlight the position and size of the NIRSpec IFS field of view.}
    \label{fig:hst_rgb_image}
\end{figure}

\aless\ is part of the Extended Chandra Deep Field South \citep{Lehmer:2005} (RA: 03h32m29.290s, DEC:-27d56m19.30s). In Figure \ref{fig:hst_rgb_image} we show the RGB image of the source.
It was identified as a strong submillimeter \citep{Coppin:2009, Debreuck:2011} and X-ray source \citep{Gilli:2011, Gilli:2014} as well as Ly$\alpha$ emitter \citep{Vanzella:2006, Vanzella:2009}. 
The presence of a narrow Ly$\alpha$, as well as a broader emission of N{\sc{v}} $\lambda$1240 {\AA} (FWHM $\sim$ 2000 \kms) \citep{Vanzella:2006, Vanzella:2009, Coppin:2009} identify the target as an AGN.
This was furthermore confirmed by the detection of X-ray emission which resulted in an estimation of the column density of $N_{\rm H} = 17.0^{+11.7}_{-6.8} \times 10^{23}$~cm$^{-2}$ \citep{Circosta:2019}  that implies the presence of a Compton thick AGN \citep{Gilli:2011, Vito:2013, Gilli:2014, Circosta:2019}. The estimated intrinsic luminosity in the 2-10 keV band is 1.3$\times$ 10$^{44}$ erg s$^{-1}$ \citep{Luo:2017} with the AGN bolometric luminosity of $2.66 \pm 0.80 \times 10^{12}~\rm L_{\odot}$ estimated via SED fitting \citep{Circosta:2019} classifying the target as a low-luminosity obscured QSO.

The kinematic of the host galaxy was extensively studied through high-angular resolution observations of the \cii 158$\mu$m emission line, revealing a massive, dusty disk with ordered rotation and low levels of turbulence (random, noncircular motion of the gas) \citep{Debreuck:2014, Lelli:2021}.
The kinematics indicate the presence of a stellar bulge, which is a sign of an already-evolved galaxy.
The fast evolution of the galaxy is also supported by metallicity measurements computed by exploiting the ratio between the FIR lines \cii158$\mu$m and \nii 205$\mu$m that report an estimated gas-phase metallicity close to solar but we note that that the measurements have large uncertainties and this tracer of metallicity is less reliable than other optical diagnostics \citep{Nagao:2012, Debreuck:2014}. 
\cite{Gilli:2014} suggest the presence of an outflow because of the observed $\sim 400$ km s$^{-1}$ velocity shift between the Ly$\alpha$ emission and other observed lines at submillimeter wavelengths (\cii, CO(2-1), [CI], CO(7-6), \nii $\lambda$205 $\mu$m; \citealt{Debreuck:2011, Nagao:2012, Damato:2020}).

\subsection{JWST observations and data reduction}

The target was observed on 12 September 2022, as part of the NIRSpec GTO ``Galaxy Assembly with NIRSpec IFS survey (GA-NIFS),'' under the program 1216 ``Integral Field Spectroscopy in GOODS-S'' (PI: Nora Lützgendorf).
The observations were carried out using a medium cycling pattern between four dithers to achieve a total exposure time of $\sim 1$h with both G235H/F170LP and G395H/F290LP gratings/filters combinations to target the emission of the galaxy from 1.7 to 5.2 $\mu$m at high spectral resolution (on average R$\sim$2700).
The raw data were downloaded from the MAST archive and then processed with a modified version of the JWST Science Calibration Pipeline version 1.8.2 with the CRDS context ``jwst\_1068.pmap.'' 
The count rate maps were created by correcting at the detector level by using the module \textit{Detector1Pipeline} of the pipeline. The calibration was performed by applying the \textit{Calwebb\_spec2} stage of the pipeline. Finally, the cube was created by adding the individual calibrated images with a \textit{drizzle} weighting and a spaxel size of 0.05\arcsec using the \textit{Calwebb\_spec3} step of the pipeline. The field of view of NIRSpec IFS is shown as  white dashed square in Figure \ref{fig:hst_rgb_image}.
Several corrections were made to the pipeline steps to allow for better data quality and to correct known bugs in the pipeline. These corrections are presented in detail in \citet{Perna:2023}, but here we mention only the major changes.
The $1/f$ correlated noise was subtracted from the count-rate images. The rejection of outliers was performed by using an algorithm similar to \verb|lacosmic| \citep{vandokkum:2001} that removes the outliers on individual 2-d exposures before combining them to create the final data cube.

\subsection{ALMA observation and data reduction}

In this work, we also use the Atacama Large Millimeter Array (ALMA) high-resolution observation of the \cii\ emission line. \aless was observed with ALMA Band-7, with a total integration time of $\sim 1.5$ h, to target the \cii\ emission line. The maximum and minimum baselines are 2516 and 14 m, respectively. 
We retrieve the \cii\ raw data from the ALMA archive (2017.1.01471.S, PI: Lelli), and then we use the pipeline scripts included in the datasets to calibrate the visibilities with the Common Astronomy Software Application \verb'CASA' \citep{Mcmullin:2007}.

Using the \verb'CASA' task \verb'tclean', we perform the cleaning on the calibrated visibilities with a \verb'natural' weighing scale and a spaxel scale of 0.05\arcsec\ to create the final datacube. The resulting datacube has a beam size of $0.17 \arcsec \times 0.14\arcsec$.

\section{Data analysis}
\label{sec:dataanalysis}

\begin{figure*}
	\includegraphics[width = \textwidth]{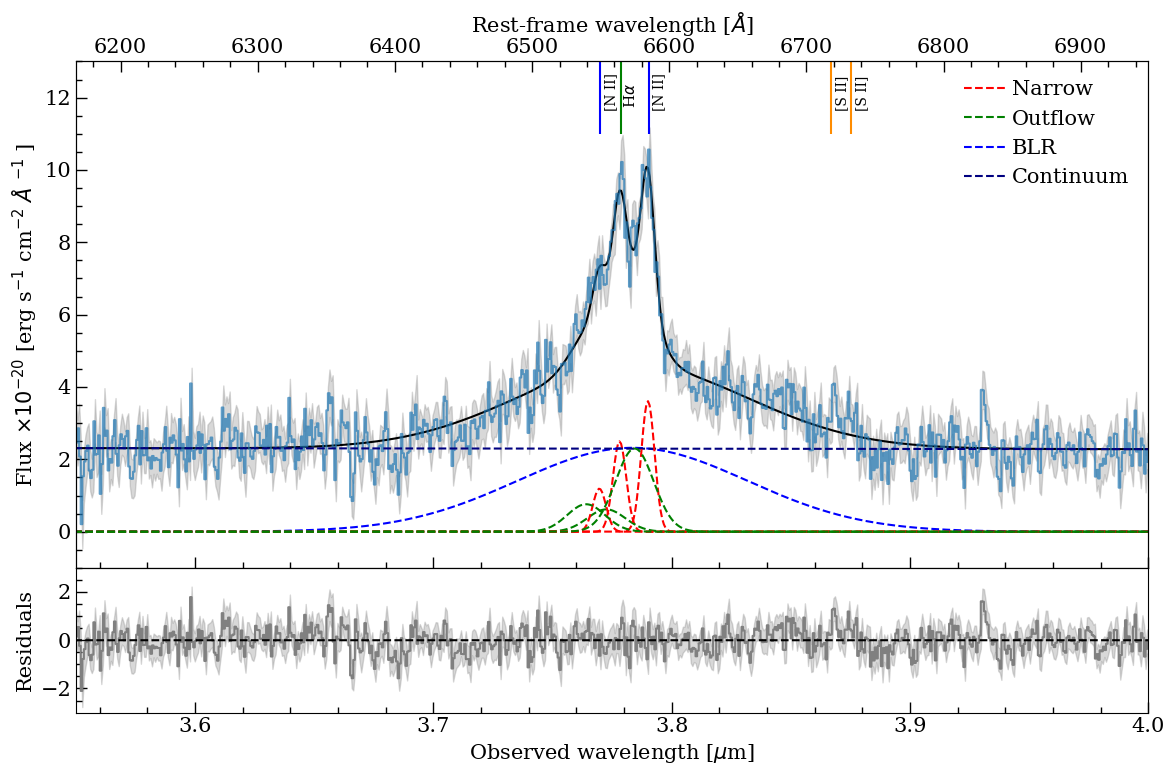}
    \caption{Spectrum in the central region from the G395H/F290LP cube. 
    In the upper panel, we report in blue the spectrum extracted from a circular aperture of radius 0.15\arcsec centered in the central region with the associated error (gray-shaded region). We show the wavelengths around the \ha\ complex. The solid black line is the best-fit model resulting in the sum of the dashed lines. 
Each dashed line represents the best-fit result of each Gaussian component or the best-fit continuum emission. In red, the emission lines associated with the narrow component tracing the host-galaxy, in green the broader component tracing the outflow, in blue the BLR, and in dark blue the best-fit polynomial continuum.
The solid vertical lines on the top represent the expected position of the \nii\, \ha\ and \sii\ lines.
    In the lower panel, we report in as a solid gray line the residuals of the fit and as a gray-shaded region the errors associated with the data.}
    \label{fig:spectrum_ha}
\end{figure*}

\begin{figure*}
	\includegraphics[width = \textwidth]{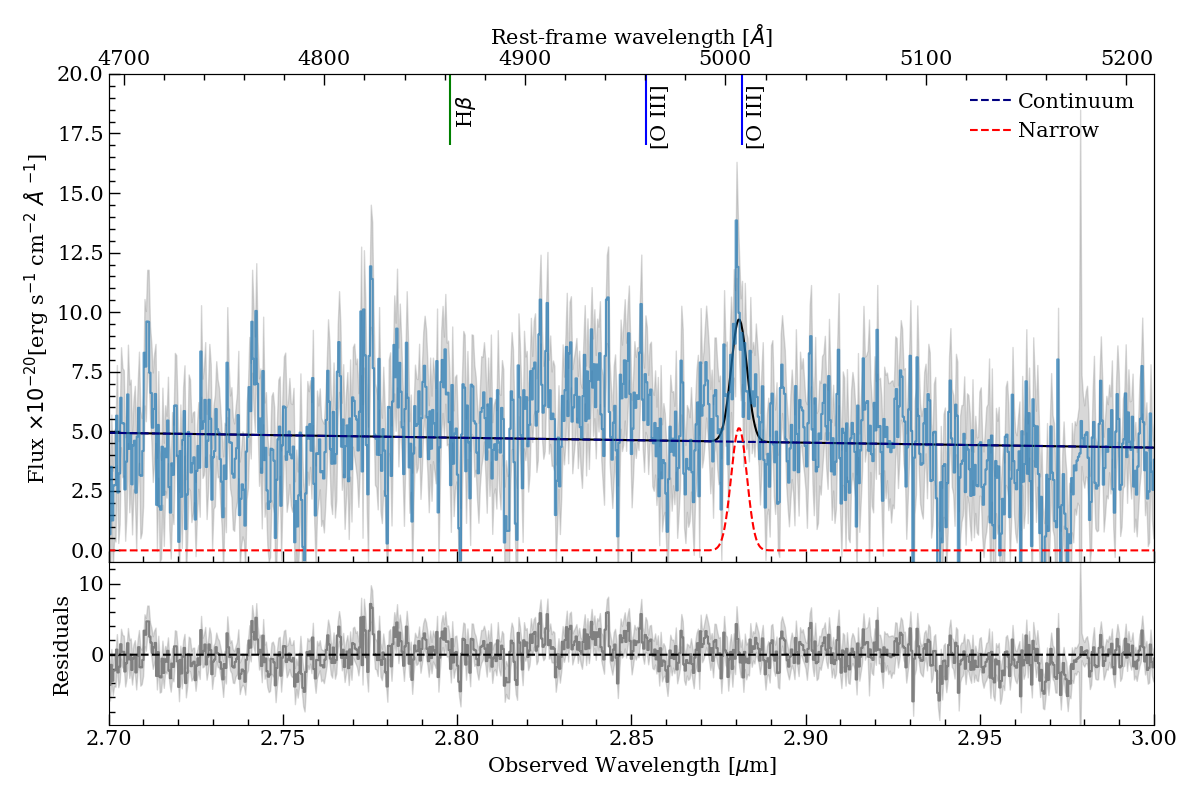}
    \caption{Spectrum of the central region from the G235H/F170LP cube. In the upper panel, we report in blue the spectrum around the \oiii $\lambda$5007\AA\ emission, extracted from a circular aperture with a radius of 0.15\arcsec\ in the central region with the associated error (gray-shaded region). In black we report the best-fit model resulting from the sum of the dashed lines. The red dashed line represents the best-fit result of the Gaussian component tracing \oiii $\lambda$5007\AA, the blue dashed line represents the best-fit continuum emission. The solid vertical lines on the top represent the expected position of the \oiii\ and  \hb\  lines.
    In the lower panel, we report in as a solid gray line the residuals of the fit and as a gray-shaded region the errors associated with the data.}
    \label{fig:spectrum_hb}
\end{figure*}

\subsection{Spectral fit of the central region}

\label{sec:spectrafitcentral}

We first analyzed the emission from the nuclear region in which we expect to find emission features associated with the hidden AGN.
Figure \ref{fig:spectrum_ha} illustrates the G395H/F290LP spectrum extracted from a circular aperture of radius 0.15\arcsec and centered on the central region. The errors on the spectrum were initially computed by summing in quadrature the noise from the error extension present in the data cube (``ERR'') in the spaxels of the selected region. To take into account the spatial correlations of the noise due to the PSF size, we scaled the errors to match the standard deviation in line-free regions of the spectrum \citep[see also][]{Ubler:2023}. 
In the spectrum, we clearly see the narrow  ($FWHM<100$~\AA) emission lines of  \ha\ and \nii\ doublets.
In addition to the narrow components, it is evident the presence of a broad ($FWHM>1000$~\AA) line associated only with the permitted line of \ha\ that is consistent with being emission from the Broad Line Region (BLR). This confirms the presence of an AGN in \aless\ and identifies the target as a Type 1 AGN, in contrast to what was expected due to the high value of $N_{\rm H}$.

The mismatch between X-ray and optical classification, despite not being so common, has been observed in both low and high-$z$ AGNs \citep{Merloni:2014, Ordovaspascual:2017, Circosta:2018, Shimizu:2018, Kamraj:2019} and it was already proposed for this target by \cite{Gilli:2014}. In particular, $\sim$ 10--23\% of optically classified Type-1 AGNs show X-ray absorbed spectra with a high value of $\rm N_H$ \citep{Scott:2012, Ordovaspascual:2017}.
This can be ascribed to different reasons: i) massive, metal-rich, high redshift galaxies can have a high ISM column density that can be large enough to absorb X-ray radiation \citep{Gilli:2014, Gilli:2022}. If the dust-to-gas ratio is low or the dust grains are large (>0.03$\mu$m), the ISM dust attenuation is not sufficiently high to    absorb UV and optical radiation resulting in an optically classified Type 1 AGN, with high X-ray obscuration \citep{Maiolino:2001b, Maiolino:2001}; 
ii) the line of sight of our observation allows us to observe the BLR unobscured from the torus, while the corona, being smaller in size than the BLR, is obscured \citep{Shimizu:2018, Kamraj:2019}.

To model the emission in the central region, we performed a least-square fitting by modeling the emission-line spectrum as a sum of Gaussian profiles. The \ha\ line profile required the addition of a broad ($ FWHM>5000$~\kms) Gaussian component to reproduce the BLR emission. The continuum emission was fitted with a power-law function. The narrow components of both \ha\ and \nii\ were modeled with two Gaussian profiles each to take into account the presence of ionized outflows because using a singular Gaussian component was not sufficient to reproduce their asymmetric profiles. To disentangle between the outflow (broad) and galaxy (narrow) components of \ha\ and \nii\ doublets,  we allowed the width of the narrow component to vary between $0<\sigma<250$ \kms, while the outflow line width was free to vary between  $250~$\kms$<\sigma<1000$~\kms. 

For each Gaussian component, we tied the centroid and line width of the \nii\ doublet to those of \ha. The two emission lines of the \nii\
doublet, originating from the same upper level, were fitted with an intensity ratio I(6584)/I(6548) fixed at $\sim$2.94 \citep{Storey:2000}. 
Finally, the model spectrum, obtained by combining all components, was convolved with a Gaussian kernel with a dispersion of $49$ \kms\ to reproduce the line spread function of the instrument at the \ha\ wavelength.  
In Figure \ref{fig:spectrum_ha} we show the best-fit profiles for each component with dashed lines, while the best-fit values for the fluxes and the FWHMs are reported in Table \ref{tab:results3by3spaxels}.

We also analyzed the spectrum from the same aperture in the G235H/F170LP cube, which covers the rest-frame wavelengths from $\sim 2900$\AA\ to $\sim 5500 $\AA. The spectrum around the \hb\ - \oiii\ complex is shown in Figure \ref{fig:spectrum_hb}. We did not identify any clear emission line, except for a tentative detection of the \oiii $\lambda$5007\AA\ with a $S/N = 2$, the low S/N of the line is possibly due to high dust extinction at bluer wavelengths. We performed a single Gaussian fitting to reproduce the \oiii $\lambda$5007\AA\ emission because the low signal-to-noise ratio (S/N) of the spectrum does not allow us to perform a multiple Gaussian fitting as performed for the \ha\ complex.
We left the width of the line free to vary between 20 and 500 \kms\ to allow for the possible presence of a broader line tracing an outflow, but the results of the fit are comparable with those obtained for the \ha\ narrow component. FWHM and flux of \oiii\ are reported in Table \ref{tab:results3by3spaxels} together with the upper limit on the \hb\ flux derived by assuming an FWHM as large as that of \ha. 

By using the ratio between the \ha\ flux and the upper limit on the  \hb\ flux, we measure a lower limit on the Balmer decrement of  $F_{\rm H\alpha, narrow + outflow}/F_{\rm H\beta} > 3.6$. This value is higher than the theoretical value for star-forming galaxies assuming case B recombination $F_{\rm H\alpha}/F_{\rm H\beta} = 2.86$ \citep{Osterbrock:2006}, implying dust absorption as expected for SMG population. Assuming a \citet{Calzetti:2000} curve we estimate a lower limit on the extinction $A_V > 0.77$ , but we expect much higher extinction as found for other SMG ($A_V \sim 4$,  \citealt{AlvarezMarquez:2023}).

To obtain a value of $A_V$, and estimate the properties of the galaxy, we performed a new SED fitting to disentangle between galaxy and AGN contribution.
The SED fitting of this target was already presented in \cite{Circosta:2019} and \cite{Gilli:2014}, but thanks to the JWST rest-frame optical observation we are now able to characterize the AGN as a type 1, hence using the correct AGN templates.
The analysis will be presented in Circosta et. al. in preparation, but here we summarize in Appendix \ref{sec:SED} the methods and report the results relevant to this work.
From the SED fitting we obtain a color excess $\rm E(B-V) = 0.55 \pm 0.07$. By using the \citet{Calzetti:2000} reddening curve, we thus calculate a dust extinction $A_V=2.2 \pm 0.3$ that is consistent with the lower limit  $A_V > 0.77$ determined by the Balmer decrement measurement.

\begin{table}
	\centering
	\caption{Results from the fitting of the spectrum extracted from a circular aperture of radius 0.15\arcsec centered on the spatial peak of the emission.}
	\label{tab:results3by3spaxels}
	\begin{tabular}{ll} 
		\hline
            \hline 

            Measurement & [km/s] 
 \\
            \hline
		$FWHM  $~\ha$_{\rm host~galaxy}$ & 519 $\pm$ 58 \\
            $FWHM  $~\ha$_{\rm outflow}$ & 1491 $\pm$ 206\\
            $FWHM  $~\ha$_{\rm BLR}$  & 9008 $\pm$ 407\\
            $FWHM  $~\oiii $\lambda$5007\AA & 553 $\pm$ 105  \\
            $\Delta v  $~\ha$_{\rm host~galaxy, outflow}$& -- 441 $\pm$ 129 \\
            $\Delta v  $~\ha$_{\rm   host~galaxy, BLR}$&  427 $\pm$ 201 \\
            \hline
            &\\
            \hline
            \hline
            Measurement & [erg/s/cm$^2$] $\times 10^{-20}$ \\
            \hline
		$F $~\ha$_{\rm host~galaxy}$ & 174 $\pm$ 34 \\
            $F  $~\ha$_{\rm outflow}$ & 123 $\pm$ 91 \\
            $F  $~\ha$_{\rm BLR}$  & 2804 $\pm$ 198 \\
            $F  $~\nii $\lambda$6584\AA $_{\rm host~galaxy}$  & 252 $\pm$ 44 \\
            $F $~\nii $\lambda$6584\AA $_{\rm outflow}$  &  460 $\pm$ 150 \\
            $F  $~\nii $\lambda$6548\AA $_{\rm host~galaxy}$  & 83 $\pm$ 14 \\
            $F $~\nii $\lambda$6548\AA $_{\rm outflow}$   & 151 $\pm$ 49 \\
            $F  $~\oiii $\lambda$5007\AA   & 291  $\pm$ 73 \\
            $F  $~\hb  & < 83  \\
            \hline

 \end{tabular}
 \tablefoot{
 On top: FWHM of the detected lines and velocity shift between the systemic velocity identified by the narrow component of the \ha\ relative to the host and the centroid of the outflow (broader component) and BLR $\Delta v  $~\ha$_{\rm host~galaxy, outflow} = v $~\ha$_{\rm host~galaxy} - v  $~\ha$_{\rm outflow}$. 
 On the bottom: fluxes of the detected lines.}
\end{table}

\subsection{Spatially resolved emission}

\label{sec:spatiallyresolved}

\begin{figure*}[ht!]

	\includegraphics[width = \textwidth]{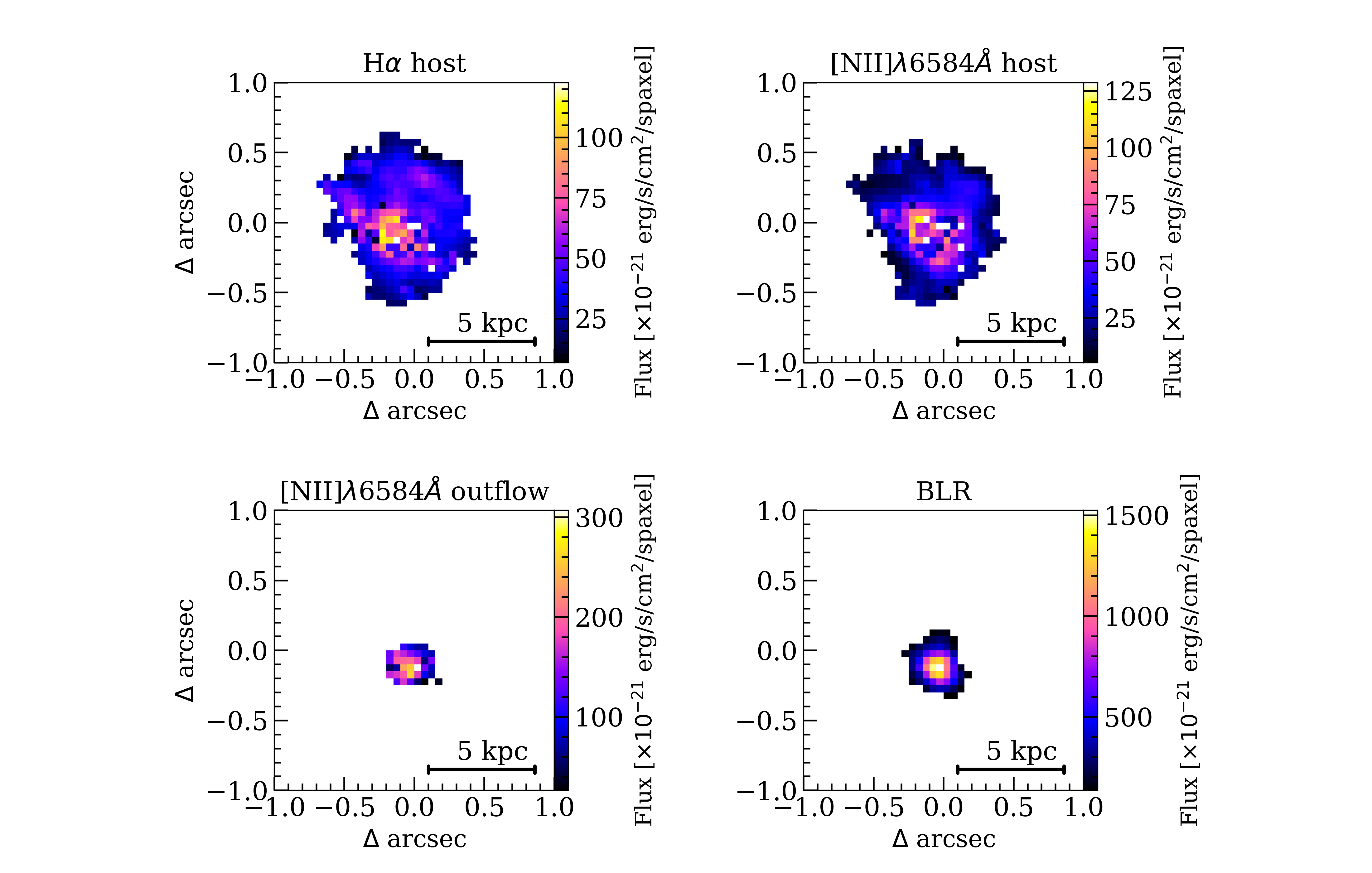}
    \caption{Flux maps created from the results of the spaxel-by-spaxel multi-Gaussian fitting. 
   Upper row: the flux of the \ha\ and \nii$\lambda$6584\AA\ narrow components representing the host galaxy emission from left to right, respectively.
    Bottom left panel: the flux of the outflow component traced by the \nii$\lambda$6584\AA\ broader component.
    Bottom right panel: the broad \ha\ component that traces the emission from the BLR. The x and y axes are the displacement in arcseconds from the galaxy center at RA = 03:32:29.3, Dec = -27:56:19.6.
    }
    \label{fig:momentmaps}
\end{figure*}

We performed a spaxel-by-spaxel fitting of the G395H/F290LP  data cube by exploiting the model adopted to reproduce the spectrum from the central region. We allowed the spectral
components to vary except for the  BLR \ha, which is  spatially unresolved, and thus its centroid and FWHM were fixed to the best-fit results obtained from the analysis of the circular aperture presented in the previous section. 

For each spaxel, two alternative models were adopted for the narrow \ha\ and \nii\ profiles: one with the outflow component and the second without it. We then selected the most suited model for each spaxel based on the Bayesian information criterion\footnote{BIC$ = \chi^2 + k \ln{N}$, where \textit{k} is the number of parameters in the fit, and N the number of data points used in the fit, we assumed Gaussian noise.} (BIC) test \citep{Liddle:2007}. For each spaxel, we estimated the BIC, and in those cases where the difference between BIC without outflow and BIC with outflow was larger than 2, we selected the model with two components as having a $\Delta$BIC$>2$ is considered marginal positive evidence in favor of the model with lower BIC value \citep{Kass:1995}.
In the other cases, we adopted the model with one component. We note that selecting the model with the lower BIC value allows us to select the best-fit model with the highest statistical significance without overfitting the data. The low S/N on a spaxel-by-spaxel analysis does not allow us to use a larger $\rm \Delta BIC$ threshold to obtain more stringent constraints on the outflow statistical significance.

Figure \ref{fig:momentmaps} illustrates the flux maps for the narrower component of \ha\ and \nii$\lambda$6584\AA\ tracing the host-galaxy,  the flux map of the \nii\ $\lambda$6584\AA\ broad component associated with outflows as it is much stronger than the one traced by \ha\ (see Figure \ref{fig:spectrum_ha} and Table \ref{tab:results3by3spaxels}) and the flux map of the BLR component.
The  \nii$\lambda$6584\AA-host  emission is predominant in the central region reaching flux values comparable with the \ha-host line (see also Sec. \ref{sec:bpt}). The \ha\ emission extends to a larger distance from the center compared to the \nii. Based on the BIC test, the additional second Gaussian profile is necessary only for the central region of the galaxy whose size is comparable to the PSF FWHM (see Appendix \ref{Appendix:PSF}). This indicates that the outflow emission is only marginally resolved by JWST and the region directly affected is limited to the central 1 kpc.

\subsection{Morphology of the host galaxy}
\label{sec:morphology}

We estimated the size of the \ha\ emission directly from the flux map obtained by collapsing the data cube in the wavelength range 3.775 -- 3.781 $\mu$m that covers the FWHM of the \ha\ narrow emission (Figure~\ref{fig:spectrum_ha}) as the flux map of the narrow component created with the pixel-by-pixel fitting (see Section \ref{sec:spatiallyresolved}) has large uncertainties due to the low S/N. 

We thus performed a two-dimensional multicomponent photometric decomposition of the map. In particular, we used a  2D Gaussian profile to reproduce the emission from the unresolved BLR, the marginally resolved outflows (see Figure \ref{fig:momentmaps}) and also taking into account the possibility of an unresolved bulge (\citealt{Lelli:2021} find that the bulge size is less than 300 pc, hence it is unresolved in our observations) and a 2D  Sérsic profile \citep{Sersic:1963} with index equal to 1 (exponential disk) to describe the emission arising from the galactic disk. We also added a 2D constant to account for a possible residual of background emission. The combination of the three models was then convolved with a Gaussian point spread function of FWHM $\simeq$ $0.202\arcsec \times 0.167$\arcsec obtained from the BLR flux map (see Appendix \ref{Appendix:PSF} for a detailed analysis).

The fit was carried out by using \textsc{Dynesty} \citep{Speagle:2020}, a Dynamic Nested Sampling Python code that allows us to estimate the Bayesian evidence and the posterior distribution of the free parameters.
For the 2D Gaussian model we assumed as free parameters the centroid position, the standard deviation along the RA and Dec directions, and the amplitude. For the 2D Sérsic model the free parameters are the normalization constant, the position of the center, the effective radius, the ellipticity, and the position angle.

\begin{figure*}
   \resizebox{\hsize}{!}
    { \includegraphics{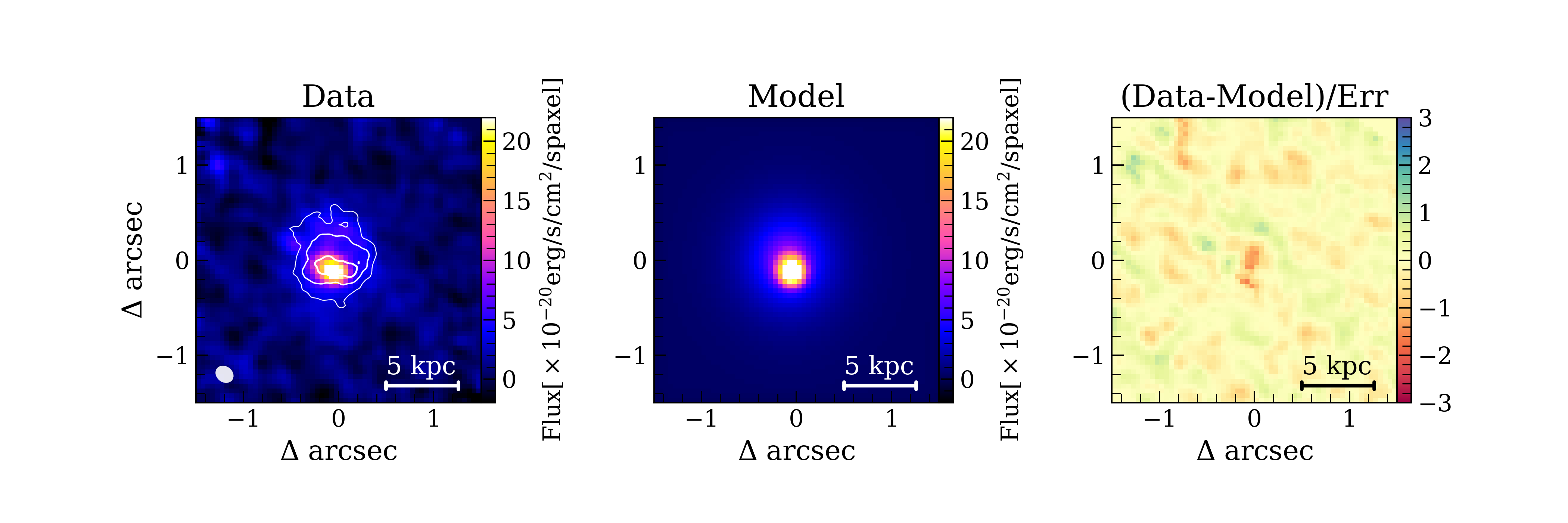}
    }
    \caption{Observed \ha\ flux map, best-fitting model, and residuals, from left to right, respectively.
    In the left panel, we show the \ha\ flux, overlaid with the \cii\ contours at 3, 6, and 9 $\sigma$. The two maps were aligned by centering them on the brightest spaxel. The gray ellipse represents the PSF size and shape. In the central panel, we show the best-fit model composed by the sum of a 2D Gaussian component that includes all the unresolved or marginally resolved components (BLR, outflow, bulge) and a 2D Sérsic component with the Sérsic index $n = 1$ that represent the galactic disk. In the right panel, we show the residuals, calculated as the observed flux minus the model, divided by the error. The color bar stretches between $-3\sigma$ and $+3 \sigma$. 
    }
    \label{fig:fluxfitting}
\end{figure*}

The data and best-fitting model are shown in Figure \ref{fig:fluxfitting}, while the corner plots for the posterior distributions of the free parameters and their best-fitting values are shown in Figure \ref{fig:corner_flux}.
The majority of the observed flux is coming from the AGN emission (BLR+outflows)  with a 2D Gaussian size of $\sim0.01$\arcsec\ that is five times smaller than the size of one spaxel and thus consistent with the emission from a point-like source. 
The effective radius of the disk is $r_{\rm e} = 0.460 \pm 0.012 \arcsec$ corresponding to an exponential scale radius of $r_{\rm D} = 1.8\pm 0.5$ kpc \footnote{$r_{\rm e} \sim 1.68 \times r_{\rm D}$}. The \ha\ disk scale radius is comparable within the errors with the disk radius measured with the \cii\ emission line \citep[$r_D = 1.4 - 1.2$ kpc][]{Debreuck:2014, Parlanti:2023}. We discuss the similarities and differences between the two tracers and the origin of such discrepancies in Section \ref{sec:discussion}.

The ellipticity of the galaxy is $0.09\pm 0.02$ corresponding to an inclination angle of $24\pm 3 \deg$\footnote{Ellipticity = $1-b/a$, where $a$ and $b$ are the semi-major and semi-minor axes, respectively. We estimate the inclination as $\cos{i} = b/a$.} assuming an infinitely thin disk. We note that the centroids of the two components have different positions (see best fit results in Figure \ref{fig:corner_flux}), with the Sérsic component being shifted northward by 0.9 kpc with respect to the BLR+outflow flux map centroid. This offset is probably caused by a central concentration of dust in the galaxy that absorbs \ha\ emission from the core \citep[see][Fig. 1a]{Lelli:2021}.

\section{Excitation mechanism}
 \label{sec:bpt}

\begin{figure}[ht!]
   \resizebox{\hsize}{!}
    {\includegraphics[width = \hsize]{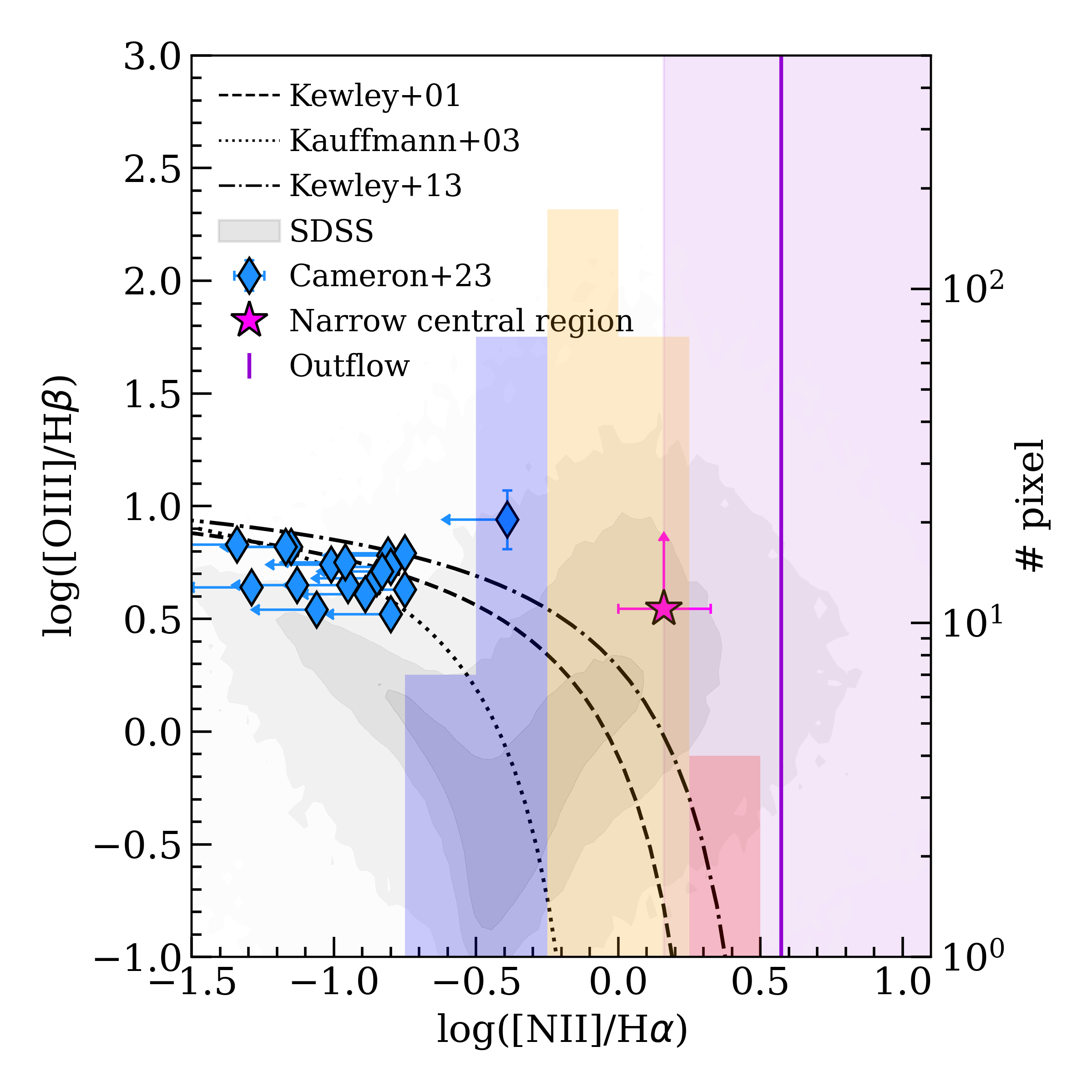}}
    \caption{BPT diagram of the target. The star marker in magenta shows the position of the central spaxels of \aless in the BPT diagram for the narrow component. The position of the outflow component based only on the \lognii\ detection is reported as a purple vertical line and the error associated with it is reported as the purples-shaded area.
    The dashed and the dotted lines are the predictions from \citet{Kewlwy:2001} and \citet{Kauffmann:2003}, respectively, for the separation between star-forming (below) and AGN (above) dominated regions at $z=0$. The dash-dotted line is the theoretical redshift evolution of the separation curve for galaxies at $z=3$ by \citet{Kewley:2013}. The gray shaded area represents the position in the BPT of SDSS galaxies  at $z \sim 0$. Blue diamonds are the results from \citet{Cameron:2023} for high-redshift ($z>5$) galaxies. 
    Overplotted with the BPT diagram we report the histogram representing the number of spaxels of the narrow component as a function of \lognii.}
    \label{fig:bpt_plot}
\end{figure}

\begin{figure*}
   \resizebox{\hsize}{!}
    {\includegraphics[width = \hsize]{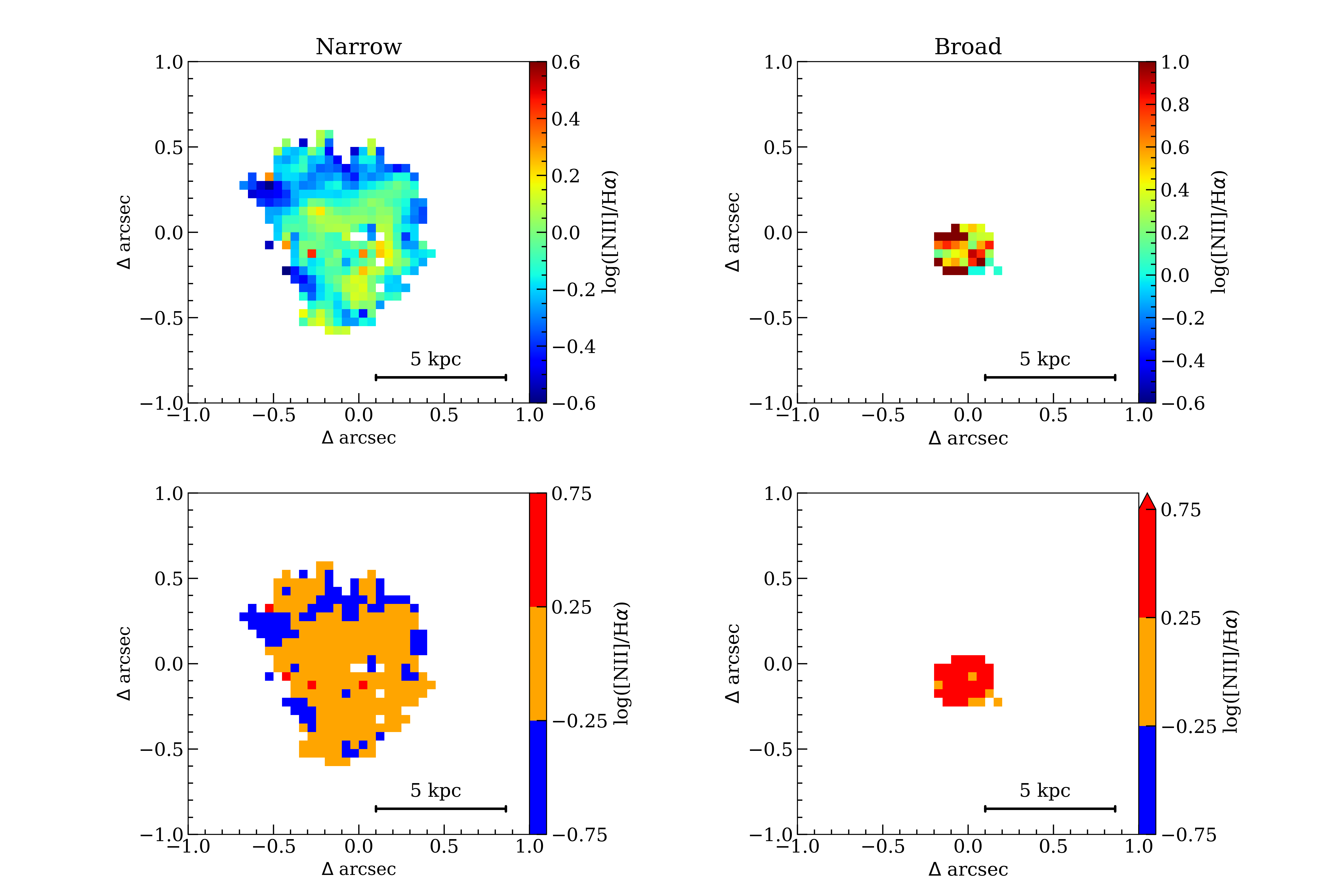}}
    \caption{Resolved BPT diagram for \aless\ host galaxy (narrow) and outflow (broad) component.
    Upper panels: spatially resolved map of the observed values of \lognii for the narrow and broad components from left to right, respectively.
    Lower panels: the spatially resolved BPT where every spaxel in the galaxy is color-coded according to its value of \lognii\ for the narrow and broad components, from left to right, respectively. as in the histogram in the upper right panel.}
    \label{fig:bpt}
\end{figure*}

The rest-frame optical emission lines can be used to characterize the primary source of excitation of the gas in the ISM of galaxies. We thus exploit the Baldwin-Phillips-Terlevich (BPT) diagram \citep{Baldwin:1981}, \oiii$\lambda$5007\AA/\hb~ versus \nii$\lambda6584\AA$/\ha, to determine the dominant source of ionizing radiation and distinguish the regions mainly excited by young stars from those where the ionization mechanism is dominated by AGN radiation. 

Figure \ref{fig:bpt_plot} shows the line ratios of the narrow component for the integrated nuclear 1D spectrum, whose line fluxes are reported in Table~\ref{tab:results3by3spaxels}.  Since we do not detect the \hb\ line, we can only report a lower limit on  \oiii$\lambda$5007\AA/\hb\ line ratio of 3.50. This is sufficient to conclude that the gas in the nuclear region of the galaxy is excited by the AGN radiation  \citep{Kauffmann:2003, Kewley:2013}. For the outflow component, we have upper limits for both \oiii\ and \hb\ and so we can report only a vertical line at the location of  \nii$\lambda6584\AA$/\ha\ line ratio in the BPT  diagram. The line ratio indicates that the photoionization by the central AGN dominates
the gas ionization with no obvious sign of a major contribution from young stars. This supports the fact that the outflowing gas is likely accelerated by AGN radiation. The kinematics of this gas will be discussed in detail in Section~\ref{sec:Outflow}.

We also investigate the spatially resolved excitation properties of the ionized gas but, since we do not detect \hb\ and \oiii\ in the individual spaxels of the data cube, we can only probe the \nii/\ha\ line ratio. In the upper left panel of Figure \ref{fig:bpt} we report the spaxel-by-spaxel value of \lognii. We observe a gradient from the central regions to the outskirts of the galaxy where the values of \lognii\ decrease at increasing radii. 
Since we only have this diagnostic to infer the excitation mechanism of the ISM in the galaxy, we define the three following possible ranges of \lognii\ based on the demarcation lines suggested by \cite{Kewlwy:2001, Kewley:2013} and \cite{Kauffmann:2003}: region likely dominated by SF excitation (\lognii~$\leq -0.25$),  region photoionized by AGN (\lognii~$  > 0.25$), and composite region $-0.25<$~\lognii~$\leq  0.25$).
In the BPT diagram in Figure \ref{fig:bpt_plot} we show the number of spaxels as a function of \lognii\ of the host galaxy component color-coded according to the aforementioned three categories, while in the lower left panel of Figure \ref{fig:bpt}  we report their spatial distribution.

The majority of the spaxels show a line ratio consistent with   ``composite''  excitation and only a few spaxels have  high enough \nii/\ha\ flux to end up in the AGN region, but they reside in the outer region of the galaxy where the S/N is lower, also they are nonadjacent, so consistent with being due to S/N fluctuations shifting the category from composite to AGN. At large radii from the center, there is a number of spaxels with low ($<0.6$) \nii/\ha\ suggesting that in these regions the excitation mechanism is likely dominated by star-formation activity. In conclusion, the spatially resolved BPT diagram indicates that the excitation mechanism is due to both an AGN and a young stellar population. Based on the results obtained from the nuclear-integrated 1D spectrum, we speculate the central part of the galaxy is mainly AGN-dominated while the excitation by young stars dominates at large radii. 
On the other hand, if we compute \nii/\ha\ spaxel-by-spaxel for the outflow component, we find that most of the spaxels have  \lognii$>0.25$ (upper and lower right panels of Figure \ref{fig:bpt}) indicating that the outflows are likely driven by AGN activity.

\section{Black hole properties}
\label{sec:blackhole}

Assuming that the gas in the BLR is virialized, we can estimate the BH mass by using the calibration by \citet{Greene:2005}:
\begin{equation}
\label{eq:BHmass}
\begin{split}
M_{\rm BH} = (2.0 ^{+0.4}_{-0.3}) \times 10^6 \left( \frac{L_{\rm H\alpha}}{\rm 10^{42}~ \rm erg~s^{-1}}\right) ^{(0.55\pm 0.02)} 
\\
 \times \left( \frac{FWHM_{\rm H\alpha}}{\rm 10^3~ \rm km~s^{-1}}\right) ^{(2.06\pm 0.06)} M_\odot
\end{split}
\end{equation}
where $ L_{\rm H\alpha}$ and $FWHM_{\rm H\alpha}$ are  the dust-corrected luminosity and the FWHM of the broad \ha\ line associated with the BLR.

\begin{figure}

	\includegraphics[width = \hsize]{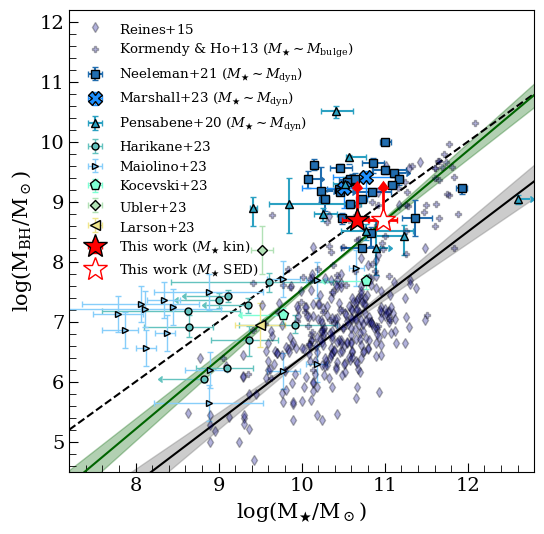}
    \caption{Relation between black hole mass and host galaxy stellar mass. The galaxy targeted in this work (\aless) is marked with a star, the star filled in red is by assuming the stellar mass derived from the kinematic fitting by \citet{Lelli:2021}, while the white filled star considers the stellar mass derived from SED fitting (see Appendix \ref{sec:SED}).
    The dark gray diamonds and crosses represent $z\sim 0$ broad line AGNs presented in \citet{Reines:2015} and the massive black holes hosted in ellipticals and spirals at  $z\sim 0$  by \citet{Kormendy:2013}, respectively.
    In light green diamond, we report the QSO at $z \sim 5.5$ studied in \citet{Ubler:2023}. In green, we show the broad line AGNs discovered with JWST at $4<z<7$ presented in \citet{Harikane:2023}, \citet{Maiolino:2023} and \citet{Kocevski:2023} as light blue circles, triangles and pentagons, respectively. The yellow triangle represents the AGN at $z \sim 8.7$ identified in \cite{Larson:2023}.
    Blue squares are the results obtained by \citet{Neeleman:2021} for QSOs at $z \sim 6$.
    Blue triangles are the results for QSOs at $z>2$ from \citet{Pensabene:2020}. Blue crosses are the results for two luminous QSOs at $z\sim 6.8$ from \citet{Marshall:2023}. We note that \citet{Pensabene:2020}, \citet{Neeleman:2021} and  \citet{Marshall:2023} report the dynamical mass rather than the stellar mass.
    The black and green solid lines are the best-fit for \citet{Reines:2015} and \citet{Kormendy:2013}, respectively, and the shaded areas are their 1$\sigma$ uncertainties.
    The black dashed line is the relation $M_{BH} = 0.01 \times M_{\star}$.
    }
    \label{fig:bhmass}
\end{figure}

We note that the lack of detection of the \hb\ line does not allow us to correct the \ha\ for the dust extinction of the galaxy and AGN torus. Hence the inferred luminosity for the BLR component is only a lower limit and consequently, we can only determine a lower limit on the BH mass: $\log{( M_{\rm BH}/M_\odot)} > 8.7$.
Figure \ref{fig:bhmass} shows the relation between the stellar mass and the black hole mass for \aless, where we use the stellar mass of  $M_\star = (4.7\pm1.6) \times 10^{10}~ \rm M_\odot$ computed by \citet{Lelli:2021} with a kinematic analysis, and $M_\star =9.5^{+4.3}_{-2.9}\times 10 ^{10} ~ \rm M_\odot$ estimated from the SED fitting (see Appendix \ref{sec:SED}).
We compare our results with those obtained from other AGN-host galaxies both at high redshift ($z>1$) \citep{Pensabene:2020, Neeleman:2020, Harikane:2023,Kocevski:2023, Larson:2023,Maiolino:2023, Ubler:2023} and in the nearby Universe ($z < 0.055$) \citep{Reines:2015}, and the local massive ``red-and-dead'' elliptical galaxies and classical bulges \citep{Kormendy:2013}. 
The estimated BH mass places \aless\ above the relation by  \citet{Reines:2015} yielding a $M_{\rm BH}/M_{\star}$ ratio $>10-30$ times higher than those estimated in local AGNs by using the stellar mass estimated by SED and kinematical fitting, respectively. However, the inferred $M_{\rm BH}/M_{\star}$ is consistent within the uncertainties with the relation determined for massive quiescent local galaxies and high redshift luminous quasars \citep{Kormendy:2013, Denicola:2019, Pensabene:2020}. As with other high-redshift AGNs and QSOs, it lies above the relation $M_{\rm BH} = 0.01 \times M_{\star}$ \citep{Decarli:2010, Denicola:2019,Pensabene:2020, Neeleman:2021}.  These high-redshift observations suggest that the BH growth dominated early on, with the galaxy catching up later. This requires that feedback and self-regulation are somehow different at early times with respect to what is observed in local AGNs.

We  can also compute a limit on the Eddington luminosity by using:

\begin{equation}
    L_{\rm  Edd} = \frac{4\pi G M_{\rm BH} m_p c}{\sigma_{\rm T}}
\end{equation}
where $m_p$ is the proton mass and $\sigma_{\rm T}$ is the Thompson scattering cross-section. We obtain an Eddington luminosity of $L_{\rm Edd} > 7 \times 10^{46}$ $\rm erg~s^{-1}$.
Comparing it with the bolometric luminosity of the AGN of $L_{\rm bol } = 1.7^{+11}_{-1.4}  \times 10^{45}~\rm erg~s^{-1}$ (see Appendix \ref{sec:SED}), we obtain an Eddington ratio of $\lambda_{\rm Edd} < 0.18 $  implying that the BH is accreting at a much lower rate than the Eddington limit. This is consistent with an evolutionary sequence where obscured AGN, like \aless, are in an early phase of QSO evolution that lasts until the Eddington ratio reaches values close to  unity and the AGN radiation is able to sweep away gas and dust from the galaxy, revealing  the emission of the bright unobscured AGNs (i.e, blue QSOs). According to this evolutionary path, we might conclude that the outflow in \aless\ is not yet energetic enough to affect the star-formation activity in the galaxy. We further investigate this possibility in Section \ref{sec:discussion}.

\section{Outflow}
\label{sec:Outflow}

\begin{figure*}
   \resizebox{\hsize}{!}
    { 
	\includegraphics{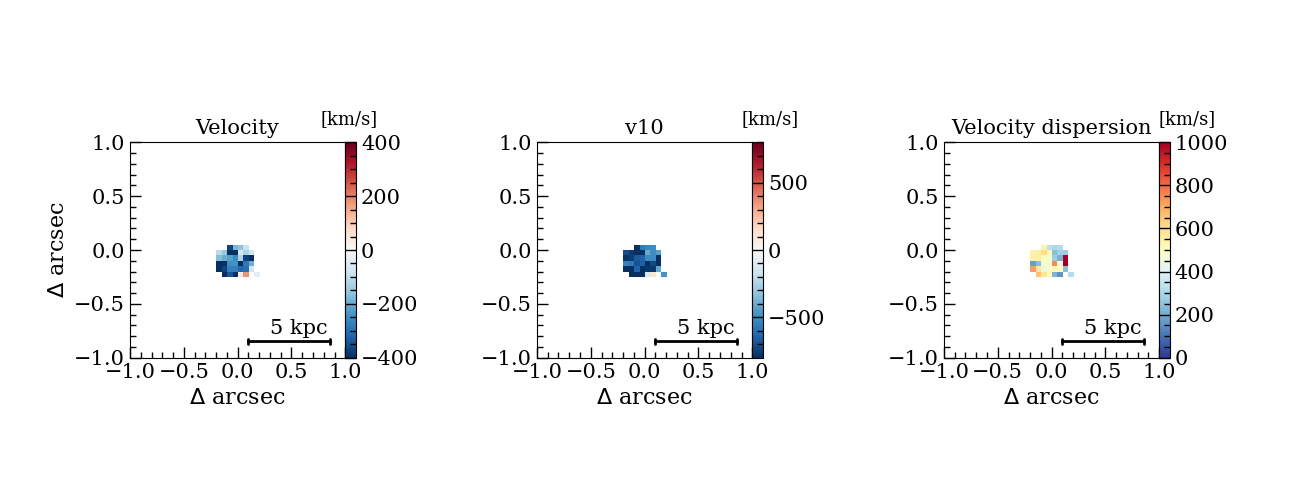}
    }
    \caption{Spatially resolved kinematic maps for the outflow component.
    From left to right, the velocity map of the outflow component, the v10 that corresponds to the velocity at which the outflow component is at the 10th percentile, and the velocity dispersion map of the outflow, respectively. Zero velocity corresponds to the velocity of the narrow line. }
    \label{fig:outflow}
\end{figure*}

In this Section, we study the mass outflow rate and the energetics of the warm ($T\sim10^4$~K) ionized gas traced by the ``outflow'' component of \ha\ and \nii\ identified as the broader component associated with each line in the Gaussian fit. 
Figure~\ref{fig:outflow} shows the kinematic maps of the outflow. The maps reveal a pattern that is not compatible with a rotating disk. Most of the spaxels in the velocity map (left panel) show negative values that are consistent with gas approaching along the line of sight. As often reported for other AGN-host galaxies \citep{Fischer:2013, Bae:2014, Perna:2017}, a corresponding redshifted component of the outflow is missing in \aless, probably due to dust obscuration of the receding side of the outflow. The central panel of Figure~\ref{fig:outflow} shows the $v_{10}$ map, the velocity at the 10th percentile of the outflow component  in each spaxel, which is usually adopted to trace the highest-velocity blueshifted gas in the outflows. We find regions in which the gas reaches a velocity as high as $v_{10} =-700$~\kms\ supporting the fact that this gas cannot be associated with the rotation of the disk, given that the maximum velocity of the rotating disk is 400~\kms\ \citep{Lelli:2021, Parlanti:2023}.

The mass of gas expelled by the outflow can be estimated by following \citet{Cresci:2023} as
\begin{equation}
    M_{\rm out} = 3.2 \times 10^5 \left(\frac{\rm L_{\rm H\alpha, outflow}}{10^{40} ~\rm erg~s^{-1}}\right) \left( \frac{100~ \rm cm^{-3}}{n_e} \right) M_\odot,
\end{equation}
where $L_{\rm H\alpha, outflow}$ is the extinction-corrected \ha\ luminosity of the outflow and $n_e$ is the electron density of the outflow. To determine the intrinsic \ha\ luminosity we correct the value in Tab.~\ref{tab:results3by3spaxels} for the value of $A_V$ estimated from the SED fitting assuming a \cite{Calzetti:2000} extinction law obtaining an extinction-corrected \ha\ luminosity of  $L_{\rm H\alpha, outflow}=  1.6 \times 10^{42} $ erg~s$^{-1}$. For the electron density, since the density-diagnostic \sii$\lambda\lambda$6716,31 line doublet is not detected in our observation, we have assumed the fiducial value of 1000 cm$^{-3}$ \citep{Forster:2019, Perna:2023, Ubler:2023} and the uncertainties are computed assuming a range of densities between 200 and 2000 cm$^{-3}$, based on the outflow densities of densities measured in high-redshift  galaxies \citep{Isobe:2023}.
We thus obtain a mass of the ionized outflow of $M_{\rm out} = 5.1 ^{+20}_{-2.5} \times 10^6 {\rm M_\odot}$.

The ionized outflow rate is calculated as follows assuming time-averaged thin expelled shells \citep{Lutz:2020}

\begin{equation}
    \Dot{M}_{\rm out} = \frac{v_{{\rm out}}M_{{\rm out}}}{R_{\rm out}}
\end{equation}
where $v_{\rm out}$ and $R_{\rm out}$ are the outflow velocity and  radius, respectively. We adopt the prescription by \cite{Genzel:2011} to estimate the velocity of the outflowing gas that takes into consideration that the emission line from the outflowing outflow is spectrally broadened due to projection effects and the velocity of the line wing traces the velocity component of the outflow directed along the line of sight, hence tracing the outflow intrinsic velocity. Thus, we obtain:  $v_{\rm out}=  |\Delta v_{\rm host~galaxy, outflow}| + 2\sigma_{\rm out} = 1710~$\kms\, where for the values of the velocity shift and the outflow velocity dispersion we use the values obtained from the fit of the spatially integrated spectrum in Sec \ref{sec:spectrafitcentral}. For the outflow extent, we use the half width at half maximum (HWHM) of the JWST PSF at 3.78~$\mu$m  given that the surface brightness emission of the outflow component is marginally resolved. We infer a mass outflow rate of $\Dot{M}_{\rm out} = 11 ^{+57}_{-5.5}~ \rm M_\odot~ yr^{-1}$.  We note that the outflow radius might be smaller than the angular resolution, therefore the reported mass outflow rate may be considered a lower limit. 

We also compute the lower limits on the kinetic and momentum rate of the ionized outflow as $\Dot{E}_{\rm out} = \frac{1}{2}\Dot{M}_{\rm out}v_{\rm out}^2$ and $\Dot{P}_{\rm out} = \Dot{M}_{\rm out}v_{\rm out}$, respectively. We obtain  $\Dot{E}_{\rm out} = 1.2\times 10^{43}\rm~erg~ s^{-1}$ and  $\Dot{P}_{\rm out} = 1.47\times 10^{35}\rm~g~cm~ s^{-2}$ with uncertainties of one order of magnitude. The outflow kinetic rate is $0.1\%$ of the bolometric luminosity. This is $\sim 50(5)$ times smaller than the theoretical values ($\Dot{E}_{\rm out}=0.05(0.005)L_{\rm bol}$) expected for quenching massive galaxies \citep{Dimatteo:2005, Choi:2012, Costa:2018, Harrison:2018}. This can suggest that the obscured AGN has not yet reached its maximum activity.

We then calculate the mass loading factor, defined as $\eta = \Dot{M}_{\rm out}/\rm SFR$, where we adopt the SFR $ = 196 \pm 10$ M$_\odot$~yr$^{-1}$ for \aless\ obtained with a SED fitting (see Appendix \ref{sec:SED}). We obtain a mass loading factor on the order of 6\%, meaning that only a minor part of the gas present in the central region is expelled compared to the gas used to create stars, and the outflow strength is probably not sufficient to remove gas and halt the vigorous, ongoing SF in the system.
We note that using other SFR estimations result in a low mass loading factor as well ($\rm SFR_{\rm [CII]}=450 \pm 70$ M$_\odot$ yr$^{-1}$, $\rm SFR_{\rm FIR}=1000 \pm 15 ~$M$_\odot$ yr$^{-1}$; \citealt{Debreuck:2014}).

\section{Gas kinematics}
\label{sec:kinematics}

\begin{figure*}
   \resizebox{\hsize}{!}
    { 
	\includegraphics{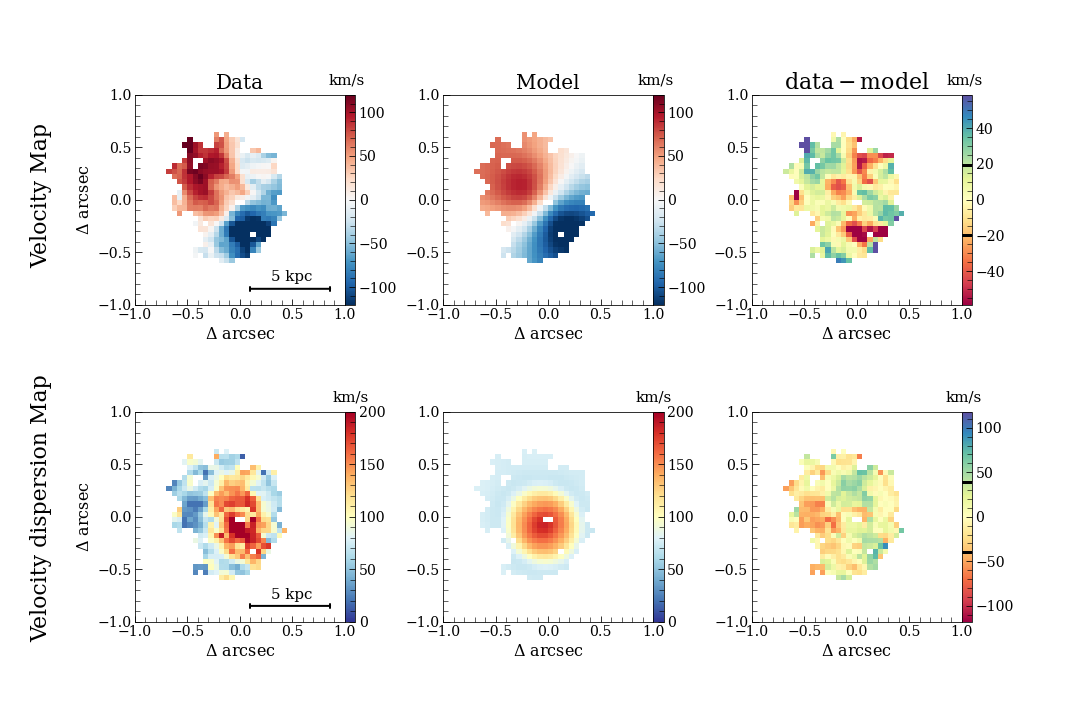}
    }
    \caption{Nonparametric best-fit results for the modeling of the \ha\ velocity and velocity dispersion maps. From left to right on the top the observed velocity map, the best-fit model map, and the residuals.
    From left to right on the bottom the observed velocity dispersion map, the best fitting model dispersion map and the residuals.
    The colorbars of the residual range
between $-3\sigma$ and $+3\sigma$, and the black lines indicate $\pm 1\sigma$.
    }
    \label{fig:maps_non_parametric}
\end{figure*}

In this Section, we investigate the gas kinematics traced by the \ha\ narrow component that maps warm ionized gas in the interstellar medium of the galaxy and we compare it with the results obtained by studying the \cii\ emission. 
The first two panels on the left of Figure  \ref{fig:maps_non_parametric} illustrate the velocity and velocity dispersion maps of the \ha\ narrow component.
The velocity and velocity dispersion maps are computed as the displacement between the centroid position of the line in that spaxel with respect to the centroid position computed in the central region, and the  standard deviation of the emission line deconvolved of the instrumental spectral resolution computed on a spaxel-by-spaxel level (see Section \ref{sec:spatiallyresolved}), respectively.
The velocity map shows a velocity gradient that spans a range of velocities between --120~\kms\ and 120~\kms\ with respect to the systemic redshift of the galaxy. The velocity pattern is consistent with that observed in \cii\ by \cite{Lelli:2021} and indicates the presence of a regularly rotating disk. We note that the velocity gradient is not symmetric on the red-shifted and blue-shifted sides, with the blue side having higher velocities closer to the center with respect to the red-shifted one, and the blue-shifted side being less extended.
On the northwestern side, we also note that the velocity pattern is irregular, similar to what is found in the \cii\ kinematic analysis \citep{Lelli:2021} that might indicate the presence of a spiral arm.

Following \cite{Parlanti:2023}, we model the \ha\ kinematics using the publicly available python library \textsc{KinMS} \citep{Davis:2013} that creates mock data cubes based on flux, velocity, and velocity dispersion radial profiles. We set up \textsc{KinMS} to simulate our NIRSpec observations, setting an angular resolution of $0.202\arcsec \times 0.167$\arcsec (see Appendix~\ref{Appendix:PSF}) and a spectral resolution of 49~\kms. We then generate the moment maps to compare directly with the observations and verify the accuracy of the model in reproducing the data. The best-fit parameter values for the kinematic model are found using the package \textsc{emcee} \citep{Foreman:2013} that allows us
to estimate the posterior probability distribution for the n-dimensional parameter space that defines our disk models by adopting a Markov Chain Monte Carlo (MCMC) algorithm.
Differently from the analysis by \cite{Parlanti:2023} we do not use a parametric function to reproduce the gas kinematics of \aless\ as the \cii\  observations reveal a complex radial profile of the velocity curve. \citet{Lelli:2021} find indeed that the de-convolved velocity is as high as $\sim400$~\kms\ in the nuclear region ($<1$~kpc) of the galaxy which is comparable with what is observed in the outskirt of the galaxy. These kinematics cannot be reproduced by the velocity curve of an exponential disk, as we show and discuss in Appendix \ref{Appendix:kin}. The flat profile of the \cii\ velocity  indeed suggests the presence of a bulge component in addition to the classic exponential kinematic component \citep{Lelli:2021, Parlanti:2023}. 

Thanks to the flexibility of \textsc{KinMS}, we adopt a nonparametric model to reproduce the observed velocity and velocity dispersion profile as a function of the radius. In particular,  the \ha\ kinematics is modeled as a series of concentric circular rings having a width of 0.6~kpc (i.e., 0.09\arcsec), which is comparable to the HWHM of the PSF of NIRSpec IFS at the \ha\ wavelength. In each ring, we assume that the emitting clouds have the same radial velocity and velocity dispersion. For the velocity fitting we assume Gaussian priors with a standard deviation of 100 \kms\ centered around the velocity of the previous ring. The velocity of the first ring was left free to vary between 0 and 1000~\kms. Similarly, for the velocity dispersion of each ring, we assume Gaussian priors with a standard deviation of 50 \kms\ centered on the velocity dispersion of the previous ring leaving the first ring free to vary between 0 and 500~\kms.
Using Gaussian priors allows us to ensure the continuity of the velocity and velocity dispersion profiles and that the discontinuities in the profiles are driven by a real increase in the likelihood. We note that assuming flat priors for each ring does not change the maps in Figure \ref{fig:maps_non_parametric}, but introduces discontinuities in the intrinsic profiles, that are then washed out by the beam smearing process when making the moment maps. 

We let the disk position angle free to vary with flat priors between 0 and 90 deg, while we fixed the inclination angle of the galaxy to 22 $\deg$ as found by \citet{Lelli:2021} and compatible within the errors with the value of inclination found in Section \ref{sec:morphology}. The emitting clouds are distributed over the rings following the surface brightness profiles obtained from the best-fit results of the flux map (Section~\ref{sec:morphology}). We note that the resolution and sensitivity do not allow us to determine stringent constraints on the velocity and velocity dispersion profiles if we adopt a disk model with a ring size smaller than the PSF HWHM due to the beam smearing. We also observe that the last ring is only probed by the outer region of the redshifted side of the galaxy as it is more extended than the blueshifted one.

The best-fit results are reported in Table~\ref{tab:npresults} and the best-fit model and residual maps of the velocity and velocity dispersion  are reported in Figure \ref{fig:maps_non_parametric}. 
In Figures \ref{fig:corner_velocity} and \ref{fig:corner_sigma} we also report the corner plot to highlight the best-fit parameters and the posterior distributions that show degeneracy between the parameters. We note that the velocity and velocity dispersion of each ring degenerate with the ones of the previous ring. In particular if one increases the other one decreases, and vice versa. This is expected and consistent with the beam smearing that acts to average out these differences.
The inferred disk position angle is in agreement with the one found for the \cii\ kinematics by \cite{Lelli:2021}.

\begin{figure*}
   \resizebox{\hsize}{!}
    { 
	\includegraphics{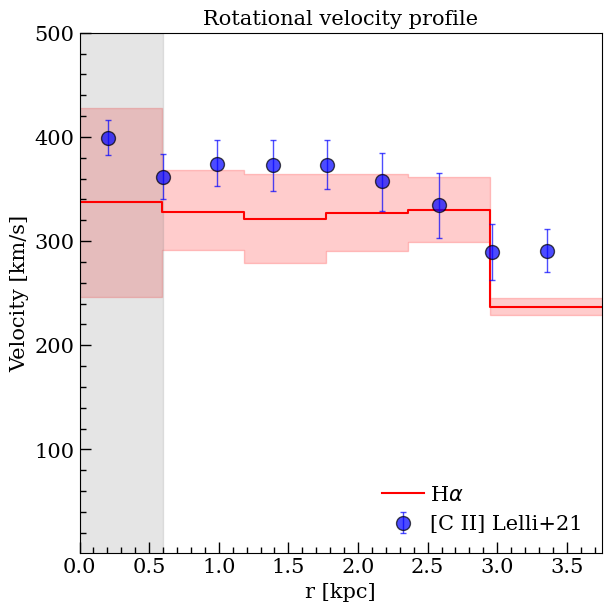}
 \includegraphics{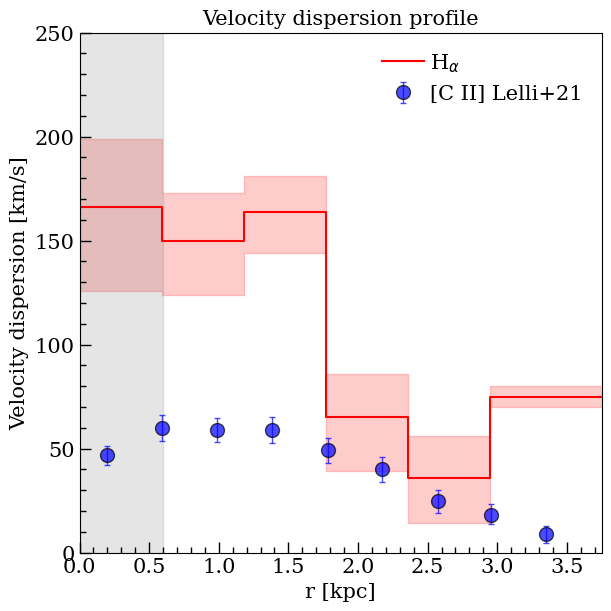}
    }
    \caption{Velocity (left) and velocity dispersion (right) profile derived from \ha\ and \cii\ from \citet{Lelli:2021}. Blue points are the results with the \cii\ tracer by \citet{Lelli:2021}. Red solid lines are the results of the nonparametric fitting of the \ha\ maps from this work with associated uncertainties. In gray is the region affected by the PSF. }
    \label{fig:vleocityprofiles}
\end{figure*}

The left panel of Figure~\ref{fig:vleocityprofiles} shows the best-fit velocity curve of \ha, which is in agreement, within the errors, with the profile inferred from \cii\ observation. The velocities reach a value of $\sim350$~\kms\ at  small scales and slightly decrease at large radii down to values on the order of $\sim250$~\kms. This result supports the scenario that \aless\ has already formed a bulge at its center that dominates the dynamics of the gas. 

The velocity dispersion profile is reported in the right panel of Figure~\ref{fig:vleocityprofiles} and spans a range between 30 and 170~\kms. Differently from the velocity curve, here we note a discrepancy between \cii\ and \ha\ gas kinematics. The velocity dispersion of the warm ionized gas is, on average, higher than the velocity dispersion of the cold gas mapped by the carbon line. In the inner 1.5~kpc the velocity dispersion observed in \ha\ is even three times larger than the one traced by \cii. 
At larger radii, the discrepancy decreases and in the two rings between 1.5 and 3.0~kpc the velocity dispersion of \cii\ is comparable (within the uncertainties) with the one derived from \ha.

As the discrepancies between our results and those found by \cite{Lelli:2021} might depend on the different approaches and tools used to fit the data, we compare directly the velocity dispersion maps in the same regions of the galaxy.
As the ALMA and NIRSpec data cubes of \aless\ have a slightly different size PSF (ALMA: 0.17\arcsec $\times$ 0.14\arcsec, JWST: 0.20\arcsec $\times$ 0.17\arcsec) we used the Python library \verb|Photutils| to create a matching kernel between ALMA and JWST PSFs. Hence we convolved the ALMA cube, which is the higher resolution one with the matching kernel to obtain 2 data cubes with the same PSF. After this process the two data cubes are affected by the same level of beam smearing, enabling us to compare the velocity dispersion maps directly.
The \cii~moment maps are created by fitting spaxel-by-spaxel singular Gaussian component to the ALMA  spectrum. We also try to include a second component to verify the presence of outflows in \cii\ but the double Gaussian model returns a higher $\chi^2$ in all spaxel and the BIC test supports the single component fit.

Figure \ref{fig:sigma_differences} illustrates the velocity dispersion of the narrow \ha\ component and \cii\ emission line with the same velocity range. We stress that the line broadening due to the line spread function of NIRSpec is corrected during the \ha\ line fitting (Section~\ref{sec:spatiallyresolved}) and the \ha\ velocity dispersion map does not include the broad blue-shifted \ha\ component associated with outflows. Comparing the kinematic maps, we find that the velocity dispersion mapped by the hydrogen line is on average higher than that traced by \cii. Both maps have high $\sigma$ at the center ($\sigma_{[CII]}$ $\sim$ 130 \kms$, \sigma_{H\alpha}$ $\sim 200$ \kms) mainly due to beam smearing, but as shown in the $\sigma_{\rm H\alpha} - \sigma_{\rm [CII]}$ map (right panel) the discrepancy between the two increases at large radii up to 0.3-0.4\arcsec. In particular, the difference in terms of velocity dispersion in the region is evident in the direction north-south from 0.1\arcsec\ to 0.4\arcsec\ from the center. In these regions $\langle\sigma_{\rm H\alpha} - \sigma_{\rm [CII]}\rangle=100$~\kms\ that is more than two times larger than the uncertainties on the velocity dispersion estimates.

\begin{table}[]
\centering  
\caption{Best fitting kinematic parameters obtained by the nonparametric fitting.}    
\label{tab:npresults} 
\begin{tabular}{cccc}

\hline\hline
                     & Radius [kpc]                                  & V [\kms]                 & $\sigma$ [\kms] \\ \hline
Ring 1               & $0\leq r<0.6 $                                    & $337^{+91}_{-91}$                            & $166^{+33}_{-40}$                  \\
Ring 2               & $0.6\leq r<1.2$                                   & $329^{+40}_{-37}$                            & $150^{+23}_{-26}$                  \\
Ring 3               & $1.2\leq r<1.8$                                   & $321^{+42}_{-42}$                            & $164^{+17}_{-20}$                  \\
Ring 4               & $1.8\leq r<2.4$                                   & $327^{+37}_{-37}$                            & $65^{+21}_{-26}$                   \\
Ring 5               & $2.4\leq r<3.0$                                   & $330^{+30}_{-31}$                            & $36^{+20}_{-22}$                   \\
Ring 6               & $3.0\leq r<3.6$                                   & $238^{+4}_{-4}$                              & $75^{+3}_{-3}$                     \\ \hline
\multicolumn{1}{l}{} & \multicolumn{1}{l}{}                              & \multicolumn{1}{l}{}                         & \multicolumn{1}{l}{}               \\ \hline \hline
\multicolumn{1}{l}{} & \multicolumn{1}{l}{$v_{\rm sys}$ [\kms]} & \multicolumn{1}{l}{$-18 \pm 1$} & \multicolumn{1}{l}{}               \\
\multicolumn{1}{l}{} & \multicolumn{1}{l}{PA [degree]}                            & \multicolumn{1}{l}{$45 \pm 1$}  & \multicolumn{1}{l}{}  
\\ \hline
\end{tabular}
\end{table}

\section{Discussion}
\label{sec:discussion}

\begin{figure*}
   \resizebox{\hsize}{!}
    { 
	\includegraphics{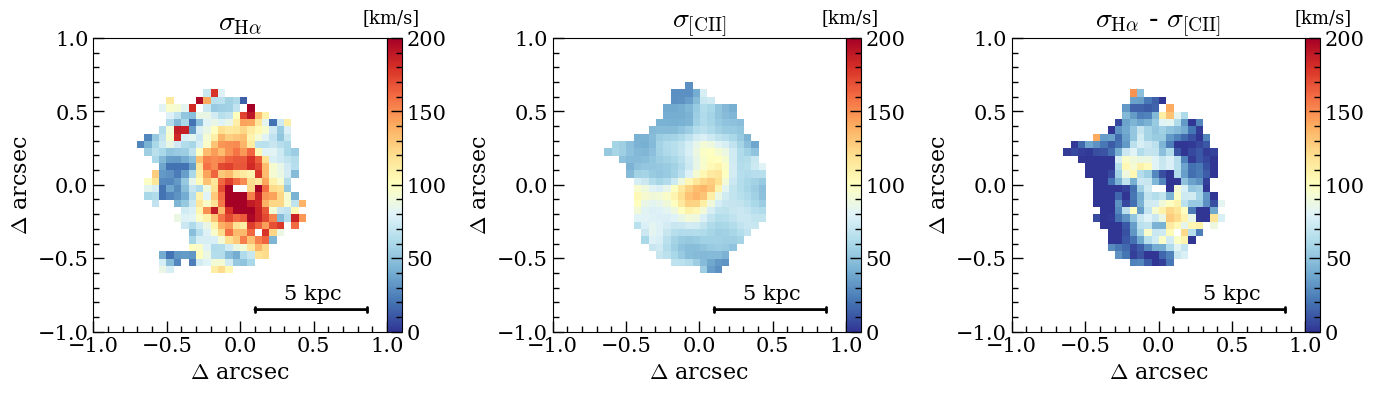}

    }
    \caption{Observed velocity dispersion maps and difference between \ha\ and \cii. In the left panel the \ha\ velocity dispersion map. In the central panel, the \cii\ velocity dispersion map rebinned to have the same spaxel size as \ha and matched to have the same PSF. In the right panel the spaxel-by-spaxel difference between the velocity dispersion of \ha\  and \cii.}
    \label{fig:sigma_differences}
\end{figure*}

Investigating galaxy dynamics at high redshifts is fundamental to understand how galaxies grow their stellar mass at their early stages of formation. Measurements of small ratios between the rotational velocity and the gas velocity dispersions are usually interpreted as evidence for turbulence in the disk due to past or ongoing strong feedback mechanisms and merging processes. Conversely, observations revealing a low velocity dispersion and $V/\sigma \sim 10$ in the gas kinematics suggest a less turbulent gas accretion and  evolution characterized by a limited number of extreme events.

The trends of velocity dispersion and the $V/\sigma$ evolution with redshift at $z>4$ are still a matter of debate today. 
On one hand, a large number of kinematic studies at $1<z<4$ suggest that high redshift galaxies are more turbulent than local ones \citep{Cresci:2009, Forster:2009, Epinat:2010, Gnerucci:2011, Ianjamasimanana:2012, Green:2014, Wisnioski:2015,  Mogotsi:2016, DiTeodoro:2016, Harrison:2017, Swinbank:2017, Turner:2017, Forster:2018, Johnson:2018, Ubler:2019, Girard:2021} with  $V/\sigma$ reaching values close to unity at $z\sim 3.5$.
On the other hand,  the results from kinematic studies at $z>4$ lead to contrasting results showing the presence of both turbulent \citep{Tsukui:2021,Herrera-Camus:2022, Parlanti:2023, Degraaff:2023} and kinematically cold galaxies \citep{Sharda:2019, Neeleman:2020, Rizzo:2020, Jones:2021, Fraternali:2021, Lelli:2021, Rizzo:2021, Posses:2023, Pope:2023} with values of $V/\sigma$ ranging from 20 to 0.1 across the redshift range $4<z<8$. However, we note that results  at high redshift are limited  either by small sample sizes for galaxies  observed with high angular resolution or by significant uncertainties due to the low angular resolution observations used for the larger samples of galaxies.

We also note that so far galaxies up to $z\sim3.5$ are principally studied by exploiting \ha\ and \oiii\ emission lines that arise from HII regions  around massive and young stars and trace the warm ionized medium with a temperature of $\sim 10^4$~K. Studies at $z>4$ target mainly the \cii\ line that principally arises from photodissociation regions (PDR) tracing the cold neutral medium at a temperature of $\sim 100$~K and only $\sim$ 30\% of its emission is associated with the warm diffuse ionized gas \citep[e.g.,][]{Stacey:1991, Stacey:2010, Croxall:2017}. Therefore, several studies have concluded that the cosmic evolution of the velocity dispersion obtained from the optical lines cannot be directly compared with that from the far-infrared \cii\ line because they are mapping different gas phases \citep{Rizzo:2022}. One possible solution is to exploit the far-infrared  \oiii$\lambda 88 \mu$m line that traces the warm ionized medium and can be observed with ALMA at $z>6$. Unfortunately, long exposure times are necessary to obtain an accurate measurement of the gas velocity dispersion, and most of the current observations have coarse angular resolution  resulting in large uncertainties \citep{Parlanti:2023}. Moreover, \oiii$\lambda 88 \mu$m would still leave a gap in observations in the redshift range $3.5 <z < 6$.

With the advent of JWST NIRSpec and in particular, thanks to the IFS mode observations, we can finally compare the kinematics determined from the rest-frame optical lines with that of \cii\ in the same $z>4$ galaxies and verify if there is a discrepancy between these tracers or not. \aless\ is the first massive ($M_\star \sim 10^{11}$ M$_\odot$) galaxy for which we have both ALMA and NIRSpec high-resolution observations and the results presented in Section \ref{sec:kinematics} show that the velocity curves of the two tracers are consistent within the errors.  The data however highlight a difference between \ha\ and \cii\ in terms of velocity dispersion. The velocity dispersion inferred from \ha\ is systemically larger by more than 50~\kms\ in the central 1.5 kpc than the one determined from the carbon line. This difference cannot be associated only with the difference in thermal velocity dispersion of the gas mapped by the two tracers as this is on the order of 20~\kms. 
The difference between the two gas kinematics is not uniform over the field of view but is  higher at the center $\langle \sigma_{\rm H\alpha} - \sigma_{\rm [CII]}\rangle= 100$ ~\kms\ and reaches values comparable with the errors ($\sim$ 30 \kms) at larger radii. Significant differences between the \ha\ and \cii\ kinematics have been also recently reported both from observations \citep{Arribas:2023} and simulations \citep{Kohandel:2023}

\begin{figure}
   \resizebox{\hsize}{!}
    { 
	\includegraphics{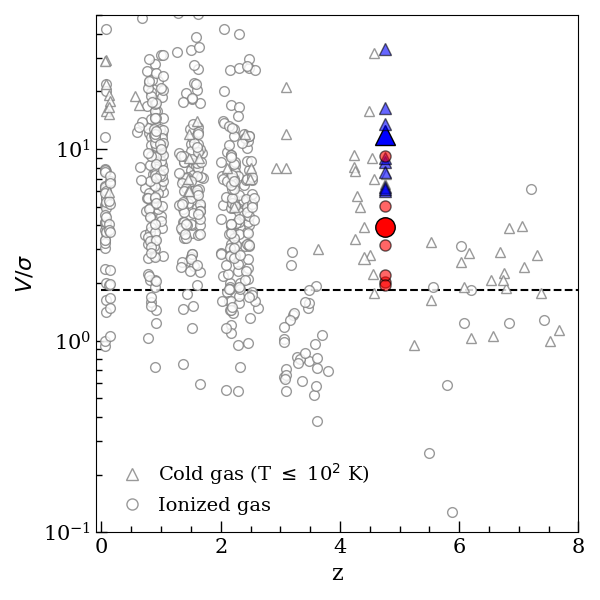}

    }
    \caption{Evolution of the ratio between the rotational velocity and the velocity dispersion with redshift. 
    Red and blue small symbols represent the value of $V/\sigma$ for each ring (see also Figure \ref{fig:vleocityprofiles}) for \ha\ and \cii, respectively. While large symbols are the mean $V/\sigma$ value across the galaxy with the two tracers.
    Gray symbols are other literature results, in particular, circles represent the results from kinematic studies that exploited ionized gas tracers (i.e., \ha, \oiii) \citep{ Green:2014,Turner:2017, Forster:2018, Wisnioski:2018, Parlanti:2023, Degraaff:2023}, while triangles represent the $V/\sigma$ values for galaxies studied though the molecular or neutral emission lines (i.e., \cii, CO) \citep{Rizzo:2020,  Fraternali:2021, Jones:2021, Girard:2021, Lelli:2021, Tsukui:2021, Rizzo:2021, Rizzo:2023}.
    The dashed line is the demarcation between rotational supported (upper) and dispersion supported (lower).
    }
    \label{fig:vsigma}
\end{figure}

To understand whether the galaxy is rotation or dispersion-supported we compute the ratio between the rotational velocity and the velocity dispersion. When the ratio is larger than $\sqrt{3.36}$ the galaxy is considered supported by rotation, on the contrary, is supported by the turbulent random motion of the gas \citep{Forster:2018}.
In Figure \ref{fig:vsigma} we report the ratio $V/\sigma$ computed for each ring for the \ha\ (red) and \cii\ (blue) as small symbols along with other high and low redshift galaxies. The big blue triangle and red circle represent the mean value of $V/\sigma$ across all the rings for the \cii\ and \ha\ profiles, respectively.
If we compare the value we obtain with the literature we see that the results with both tracers lie in the rotation-dominated region, even if the \ha\ and \cii\ points create two different clouds around the central value of $V/\sigma = 10$ for the \cii, and $V/\sigma = 3.5$ for the \ha. 
The ratio between the rotational velocity and the velocity dispersion derived from the \ha\ line is $V/\sigma \sim 10$ in the outer regions as also found with the \cii\ tracer for this galaxy \citep{Lelli:2021} and other high-redshift dusty star-forming galaxies \citep[e.g.,][]{Rizzo:2020, Rizzo:2021, Fraternali:2021}, and decreases down to values of $\sim 2$ in the central region, consistent with a thick turbulent rotating disk.

The enhanced velocity dispersion of the warm ionized gas can be caused by the outflows driven by the central AGN. Theoretical models predict that outflows might remove gas from the galaxy as well as inject energy into the interstellar medium and kinematically heat the gas. The increase in the turbulence in the ionized phases of the ISM due to the feedback effect has been recently observed in \citet{Marasco:2023} at $z = 0$, while a correlation between the increase of the turbulence in the galaxy and the presence of a central AGN that can power a nuclear outflow has been found in \citet{Ubler:2019}.
The \ha\ gas might be  affected by the galactic winds and its kinematics reflects the impact of the outflows on the host galaxy. We thus compare the kinetic energy of the gas with that injected by the outflows into the galaxy.
We estimate the  kinetic energy of the warm gas traced by the narrow \ha\ component, $E_{\rm H\alpha}= M_{\rm H\alpha}\sigma_{\rm H\alpha}^2/2=1 \times 10^{54} ~{\rm erg}$, by employing the \ha\ flux and velocity dispersion in the region with $\sigma_{\rm H\alpha} - \sigma_{\rm [CII]} > 70$~\kms\, that is the region where the difference between the velocity dispersions in the third panel of  Figure~\ref{fig:sigma_differences} is two times larger than the median error.
 We compare the inferred  kinetic energies with the outflow energy, $E_{\rm out}= \langle \Dot{E}_{\rm out} \rangle \tau_{\rm out}=2.3 \times 10^{57}$ erg, where we assume that the outflow kinetic rate is constant over time and the feedback mechanism started  10 Myr ago ($\tau_{\rm out} = 10$ Myr). 
 We find that the ratio between the energy provided by the outflow, and the energy necessary to increase the turbulence of the ISM is $E_{\rm out}/E_{\rm H\alpha} = 0.05\%$.
 The result suggests that the energy of the outflows is powerful enough to provide the kinetic energy of the ionized gas observed in the line-width enhancement region.
In the outer regions of the galaxy, instead, the warm and cold gas are coupled as the velocity dispersion of both tracers reaches values of 30~\kms\ and we do not find evidence of the feedback from the weak outflow or the accreting BH enhancing the turbulence at larger radii. The level of turbulence in the outskirts of the galaxy can easily be sustained by star-formation feedback, gravitational instabilities due to the accretion of gas on the disk or due to the transport of gas from outer to inner radii \citep{Krumholz:2016, Krumholz:2018, Ginzburg:2022}.

\section{Conclusions}
\label{sec:conclusion}

In this work, we have presented the JWST/NIRSpec Integral Field Spectrograph (IFS)
observation of the 
AGN-host galaxy \aless\ at $z = 4.755$. The observations of the high-resolution gratings have allowed us to study for the first time the rest-frame  optical emission lines of a dusty-obscured SMG hosting an AGN. In particular, we exploited the \ha\ and \nii\ emission lines to trace the host galaxy kinematics, determine the presence of  a BLR, and investigate the properties of ionized outflows. 
Our main results are the following:

\begin{itemize}
    \item We observe a broad \ha\ component with a FWHM of $\sim$ 9000 \kms\ arising from the BLR around the accreting supermassive black hole. The presence of the BLR unambiguously classifies the target as a type 1 AGN, in contrast to what was believed before due to the high observed column density of $N_{\rm H}\sim10^{24}~{\rm cm^{-2}}$. The broad line emission implies a BH mass of $\log(M_{\rm BH}/M_\odot) > 8.7$ that is slowly accreting at a smaller rate than the Eddington rate ($\lambda_{\rm Edd}$<0.18).

    \item On the $M_{\rm BH} - M_\star$ plane, the target lies on the relation for massive quiescent local ellipticals, classic bulges, and luminous QSO at high redshift. But $M_{\rm BH} /M_\star$ is more than one order of magnitude higher with respect $M_{\rm BH} /M_\star$ observed in local AGN with similar stellar masses.

    \item We find hints of a weak, marginally resolved, ionized outflow with a mass loading factor of $\sim $0.06, implying that the outflow is not able to eject away a large amount of gas to halt the SF ongoing in the galaxy. 

    \item By measuring the ratio between \nii\ and \ha\ we have found that the AGN hard radiation is the dominant source of ionization of the ISM, especially in the central region of the galaxy and for the outflow component. At larger radii we find, instead, softer radiation, compatible with emission from young, bright stars.

    \item Despite the low mass loading factor, the ionized outflow seems to be sufficiently energetic to increase the turbulence in the system.  In fact, the kinematic analysis of the \ha\ line shows that gas turbulence in the central region is 2-3 times higher than the rest of the galaxy. This increase in turbulence might be the initial effect of the outflow on the host galaxy. However, as such high-velocity dispersion is observed only in the \ha\ kinematics and not in the map of \cii, we conclude that the outflow is injecting turbulence in the warm and diffuse ionized gas, but it is not sufficiently powerful to disrupt the dense gas and
    quench star formation.

\end{itemize}

The complex scenario of galaxy-black hole coevolution is still far from being constrained and firmly established. 
In this work, we have highlighted how JWST with its high spatial resolution, spatially resolved spectroscopy capability, and an infrared wavelength range probed with unprecedented sensitivity, significantly enhance previous studies based on observations at various wavelengths.
In particular, its ability to probe the rest frame optical emission lines at high redshift with high spectral and spatial resolution has allowed us to study the first phases of the interplay between the accreting black hole and the host galaxy and connect kinematic measurements of high-$z$ tracers (\cii) with the more traditional rest-frame optical emission line tracers at lower redshift.
Future JWST IFS observations, alongside other ground-based facilities (e.g., ALMA) will allow us to better observe and understand the phenomena involved in the complex environment of dusty massive galaxies in the early Universe and their interaction with the massive black holes they host. \aless\ observations suggest that SMGs could be the evolutionary stage preceding the active QSO phase. The accreting BHs in SMG have not yet reached the Eddington limit and outflows are not powerful enough to remove gas from the galaxy, but they are injecting energy into the system and increasing the turbulence of less dense gas.

\begin{acknowledgements}
We thank the anonymous referee for providing us constructive comments and useful suggestions, Fabio Vito for useful comments and discussions, and Emily Wisnioski for sharing v/$\sigma$ measurements from the KMOS$^{3D}$ survey \cite{Wisnioski:2018}.
SC and GV acknowledge support from the European Union (ERC, WINGS,101040227).
FDE, RM and JS and acknowledge support by the Science and Technology Facilities Council (STFC), by the ERC Advanced Grant 695671 ``QUENCH,'' and by the UKRI Frontier Research grant RISEandFALL. RM is further supported by a research professorship from the Royal Society.
SA, BRdP, MP acknowledge grant PID2021-127718NB-I00 funded by the Spanish Ministry of Science and Innovation/State Agency of Research (MICIN/AEI/ 10.13039/501100011033).
MP acknowledges the Programa Atracci\'on de Talento de la Comunidad de Madrid via grant 2018-T2/TIC-11715.
IL acknowledges support from PID2022-140483NB-C22 funded by AEI 10.13039/501100011033 and BDC 20221289 funded by MCIN by the Recovery, Transformation and Resilience Plan from the Spanish State, and by NextGenerationEU from the European Union through the Recovery and Resilience Facility.
GC acknowledges the support of the INAF Large Grant 2022 "The metal circle: a new sharp view of the baryon
cycle up to Cosmic Dawn with the latest generation IFU facilities."
AJB, AJC, GCJ and JC acknowledge funding from the "FirstGalaxies" Advanced Grant from the European Research Council (ERC) under the European Union’s Horizon 2020 research and innovation programme (Grant agreement No. 789056).
H{\"U} gratefully acknowledges support by the Isaac Newton Trust and by the Kavli Foundation through a Newton-Kavli Junior Fellowship.
This paper makes use of the following ALMA data: ADS/JAO.ALMA\#2017.1.01471.S ALMA is a partnership of ESO (representing its member states), NSF (USA) and NINS (Japan), together with NRC (Canada), MOST and ASIAA (Taiwan), and KASI (Republic of Korea), in cooperation with the Republic of Chile. The Joint ALMA Observatory is operated by ESO, AUI/NRAO and NAOJ.
   
\end{acknowledgements}

\bibliographystyle{aa}
\bibliography{aa}

\begin{appendix} 
 \section{SED fitting}
\label{sec:SED}
To decompose the contribution from the AGN and the host galaxy and infer quantities that will be useful in our analysis we performed the SED fitting by using the publicly available code \verb|CIGALE|\footnote{https://cigale.lam.fr/} (Code Investigating GALaxy Emission) \citep{ Boquien:2019}. For the fitting, we followed a procedure similar to the one used in \cite{Circosta:2018}. The SED fitting for the whole GA-NIFS sample will be presented in Circosta et al. in preparation, here we summarize the method, and we report the main results for this target.

We used all the available photometry in the literature from the rest-frame UV wavelengths to the FIR. The UV-to-NIR photometry is taken from \cite{Merlin:2021}, while \textit{Spitzer}/MIPS, \textit{Herschel}/PACS, and \textit{Herschel}/SPIRE data are from the catalog presented in \cite{Shirley:2021}.
Datapoints with S/N $>$ 3 were considered detections, or converted to 3$\sigma$ upper limits otherwise.
The rest-frame UV, optical, and NIR SED are well sampled by \textit{HST}, Subaru, and \textit{Spitzer}/IRAC data, while in the MIR we have only an upper limit from \textit{Spitzer}/MIPS 24$\mu$m filter.
Our JWST observations allow us to probe the wavelength range from 1.7 to 5.3 $\mu$m allowing us to compare the fluxes we estimate with the available IRAC photometry. By convolving our observation with the 3.6$\mu$m and 4.5$\mu$m IRAC filter response we obtain values comparable to the one in the literature, hence we decided not to include these data in the SED fitting.
The far infrared part of the SED is constrained by upper limits from \textit{Herschel}/PACS, and \textit{Herschel}/SPIRE, and the continuum emission in ALMA bands 4, 6, and 7, from the ALMA archive (PID: 2015.1.00040.S), \cite{Debreuck:2014}, and \cite{Nagao:2012}, respectively.

By using a Bayesian approach the code finds the best SED model by combining the light emitted from stars, dust heated by SF, and the AGN, taking into account the energy balance between the light absorbed in the UV-optical and reemitted in the FIR.
The stellar emission is reproduced by assuming a delayed exponential star formation history plus a recent burst, a Chambrier IMF \citep{Chabrier:2003}, and stellar population models by \cite{Bruzual:2003} with solar metallicity. The emission is then attenuated by dust by using a modified Calzetti attenuation law \citep{Calzetti:2000}. Dust emission heated by star formation is reproduced by the models presented by \cite{Dale:2014}.
As for the AGN contribution, we used the models by \cite{Fritz:2006}, and we used a type 1 AGN template.

\begin{figure}[h!]
    \resizebox{\hsize}{!}
     { 
 	\includegraphics{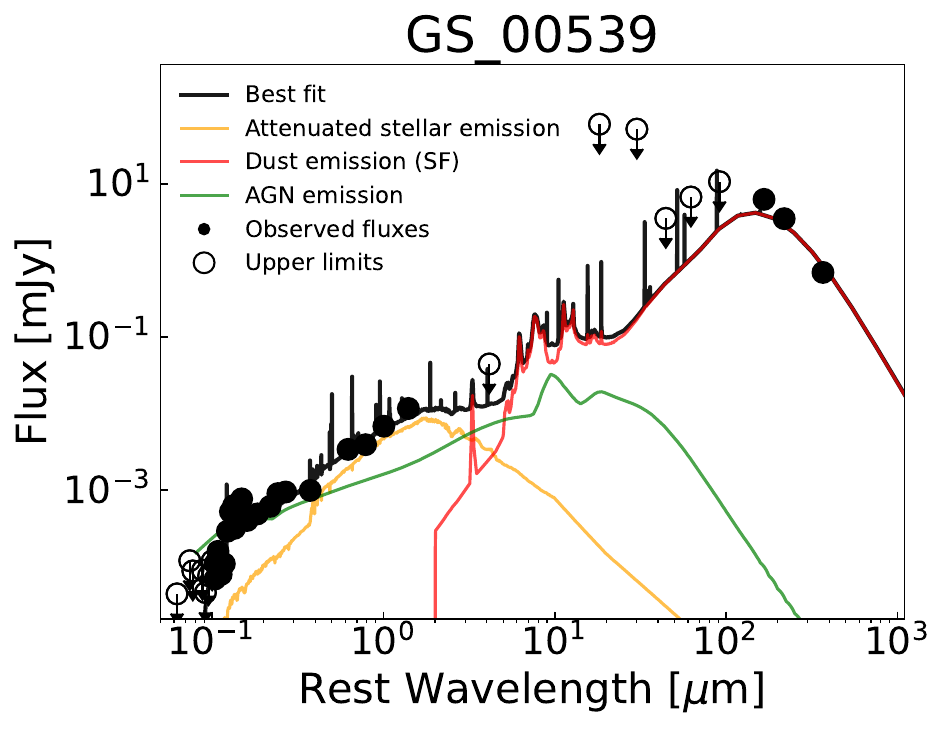}
     }
     \caption{SED fitting of \aless\ in rest wavelengths (Circosta et al. in prep.). The black circles are all the available photometric points, while the white-filled circles represent the $3\sigma$ upper limits.
     The black solid line is the best-fit result.
     The yellow, red, and green solid lines are the best models for the attenuated stellar emission, dust emission, and AGN emission, respectively.}
     \label{fig:sed}
 \end{figure}

The result SED is shown in Figure \ref{fig:sed} and the best-fit values are presented in Table \ref{tab:SED_results}.

\begin{table}[h!]
    \centering
        \caption{SED fitting best-fit results.}
    \begin{tabular}{ll}
    \hline
    \hline
 Measurement&Best-fit value\\
     \hline

         $\log{(\rm M_\star/M_\odot)}$& 10.98 $\pm$ 0.16 \\
         SFR [$\rm M_\odot~yr^{-1}$]&  196 $\pm$ 10\\
 $\log{\rm (L_{\rm bol}/[erg~s^{-1}])}$& 45.22 $\pm$ 0.88\\
 E(B-V)& 0.51 $\pm$ 0.07\\
 \hline
 \end{tabular}

    \label{tab:SED_results}
\end{table}

 \section{IFS point spread function}
 \label{Appendix:PSF}
 The fitting of the moment maps to recover the galaxy size and kinematic parameters requires adequate modeling of the telescope point spread function (PSF) that allows us to correct and take into account the beam smearing effect.
 The beam smearing is introduced when the light from a point in the sky plane is spread over a larger area on the detector.
 This effect has a consequence both in the spatial and the spectral axis due to the reconstruction of the final cubes. In particular, on the moment maps the flux map will have a larger extension, the velocity map will have a shallower velocity gradient, and emission lines will be artificially broadened resulting in a larger velocity dispersion, especially in the central region of the galaxy.
To correct these artifacts and recover the intrinsic model parameters we convolve our kinematic model with the telescope PSF, in that way, we can directly compare the model and observations.

The \verb'webbpsf' python tool \citep{Perrin:2014} generates the theoretical PSF by taking into account instrumental properties and the optics based on test results. Since JWST has wavelength-dependent PSF (see also \citealt{Deugenio:2023}) we simulated the PSF at the wavelength at which we observe the \ha\ line (3.778 $\mu$m). The generated PSF presents spikes and irregularities and fitting it with a 2D Gaussian we obtain a FWHM of $0.11 \times 0.10$\arcsec.

\begin{figure*}
    \resizebox{\hsize}{!}
     { 
 	\includegraphics{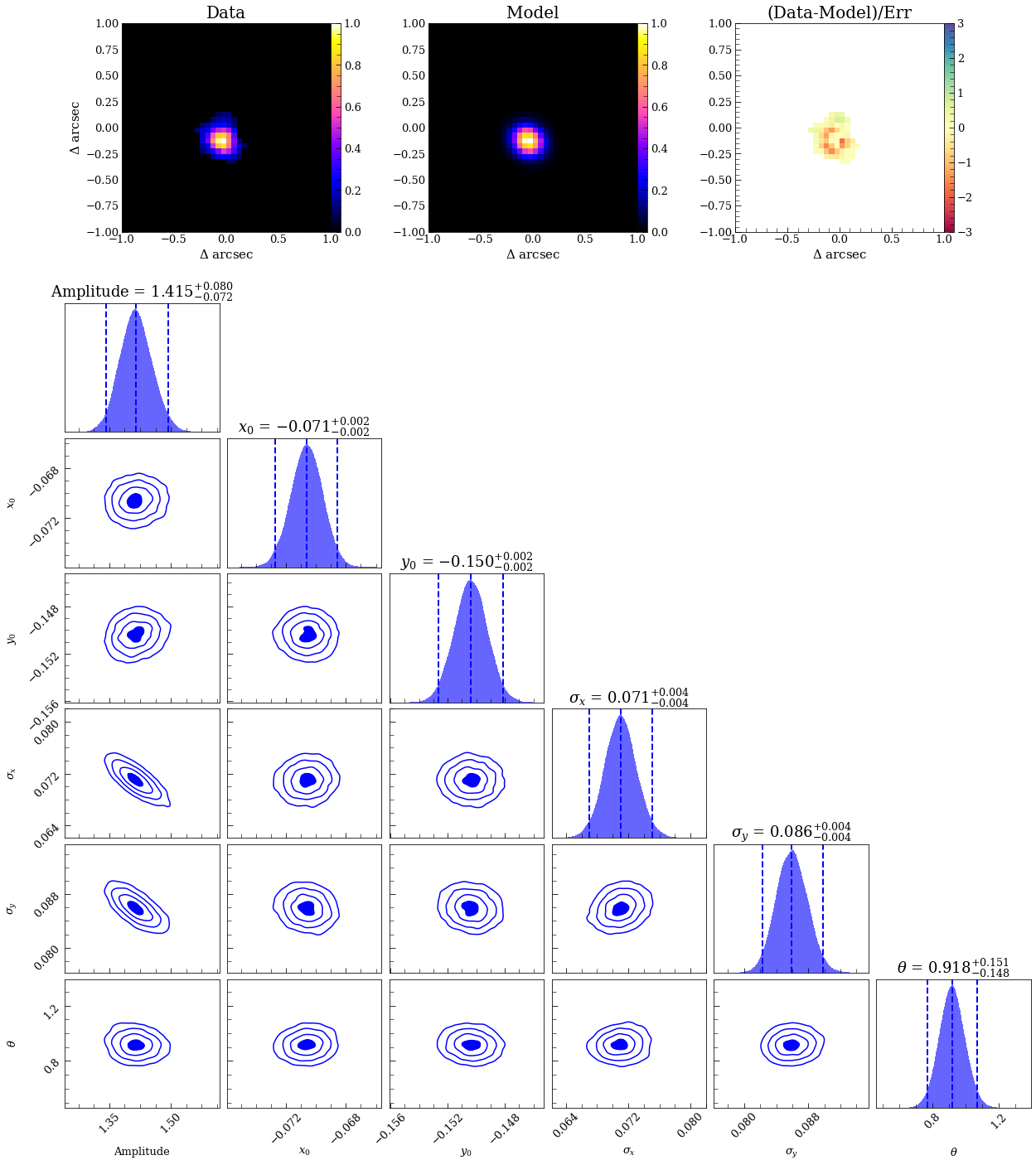}
     }
     \caption{Results of the Gaussian fitting of the BLR to recover the observed point spread function. On the top, the BLR observed flux maps, the best-fit model, and residual from left to right. On the bottom, the corner plot reports the best-fit results and uncertainties.
     The amplitude is a normalization constant, $x_0$ and $y_0$ are the  displacement in arcsec of the centroid from the galaxy center coordinates, $\sigma_x$ and $\sigma_y$ are the minor and major axis standard deviations in arcsec, $\theta$ is the rotation angle in radians measured counterclockwise from the positive y-axis.
     Above each histogram, we show the best value for each parameter and the 2$\sigma$ interval.}
     \label{fig:BLR_PSF}
 \end{figure*}

Since \verb'webbpsf' only simulates the NIRSpec PSF if used in imaging mode without taking into account the IFS optics, the dithering schemes, and the cube reconstruction processes, we also have fitted with a Gaussian function the BLR flux map. The broad line emission arises from a region close to the black hole \citep{Wandel:1999}, hence it will not be spatially resolved and its shape should reflect the instrumental PSF. The results of the fitting are shown in  Figure \ref{fig:BLR_PSF}. The results of the fitting provide an elliptical PSF with a FWHM of $0.202\arcsec \times 0.167$\arcsec where the major axis is rotated counterclockwise in respect to the x-y axis of an angle $\theta = 52 \pm 4 \deg$. The resulting PSF is elongated in the cross-dispersion direction (i.e., slices direction).
As the BLR emission should reflect the real telescope PSF we assumed the BLR best fitting results as our fiducial model for the PSF to use in our kinematic analysis.

 \section{Morphological and kinematic fitting}

\label{Appendix:kin}

In this Section, we present the outputs and corner plot for the morphological and kinematic fitting discussed in Sections \ref{sec:morphology} and \ref{sec:kinematics} and we also present an alternative kinematic fitting including only an exponential disk to show that the nonparametric approach is the best model to reproduce this galaxy kinematics.

\subsection{Posterior distributions}

\begin{figure*}
    \resizebox{\hsize}{!}
     { 
 	\includegraphics{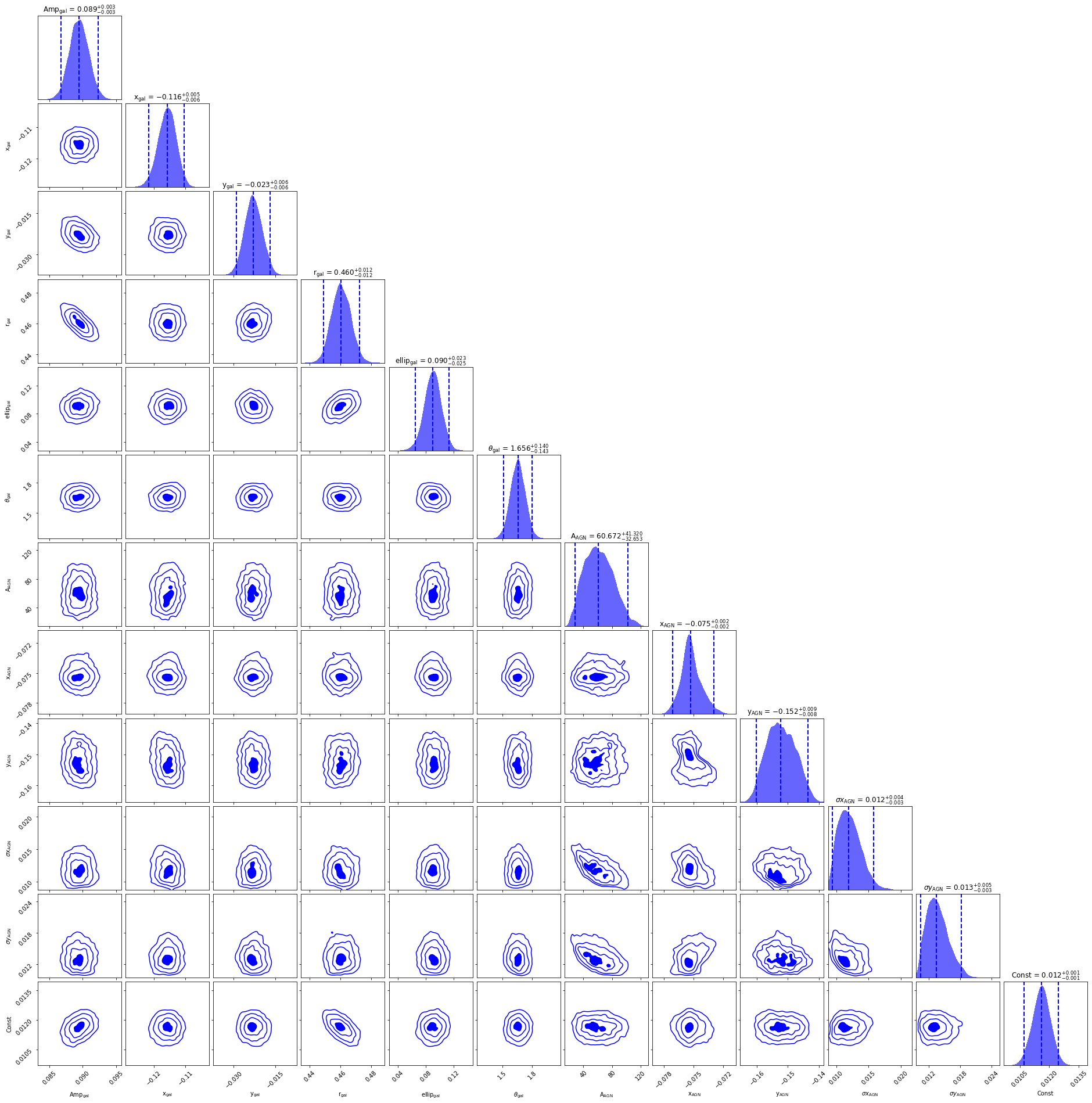}
     }
 \caption{Posterior distribution for the free parameters in the fitting of the flux map. The dashed lines represent the 16th, 50th and 84th percentile. The galaxy is modeled as a 2D Sérsic profile, while the AGN is modeled as a two-dimensional Gaussian.
     The coordinates of the centers are expressed in arcseconds where the position (0,0) corresponds to the coordinates of RA = 03:32:29.3, Dec = -27:56:19.6.
     The radius of the galaxy and the standard deviation of the Gaussian component $\sigma_x$ and $\sigma_y$ are measured in arcseconds and  
     $\theta_{\rm gal}$ is the angle measured in radians between the y-axis and the semi-major axis of the galaxy.
     }
     \label{fig:corner_flux}
 \end{figure*}

\begin{figure*}[ht!]
    \resizebox{\hsize}{!}
     { 
 	\includegraphics{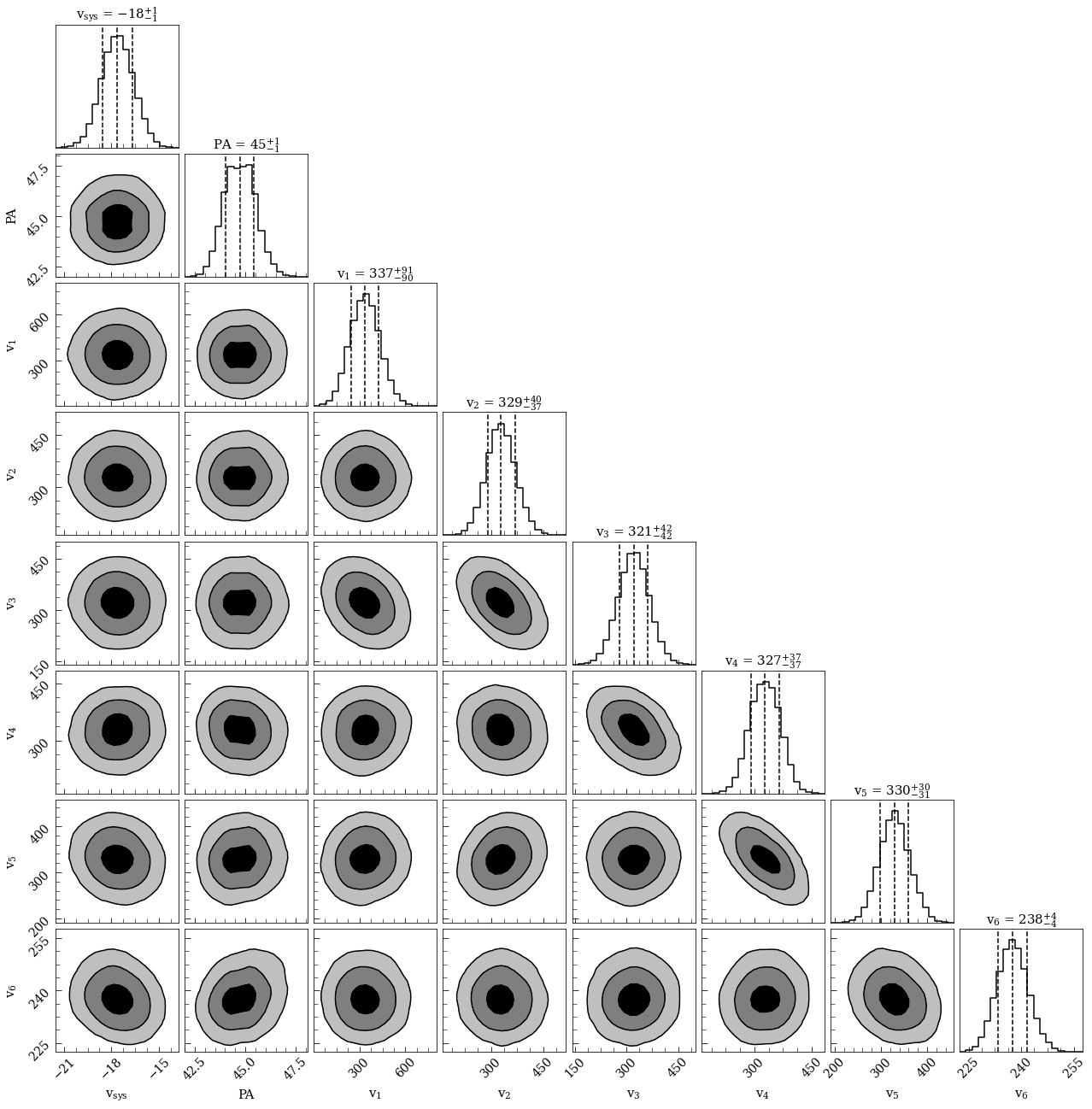}
     }
     \caption{Posterior distribution for the free parameters in the velocity map fitting. The dashed lines represent the 16th, 50th and 84th percentile. The units of the velocities are \kms, the PA is measured in degrees.}
     \label{fig:corner_velocity}
 \end{figure*}

\begin{figure*}[ht!]
    \resizebox{\hsize}{!}
     { 
 	\includegraphics{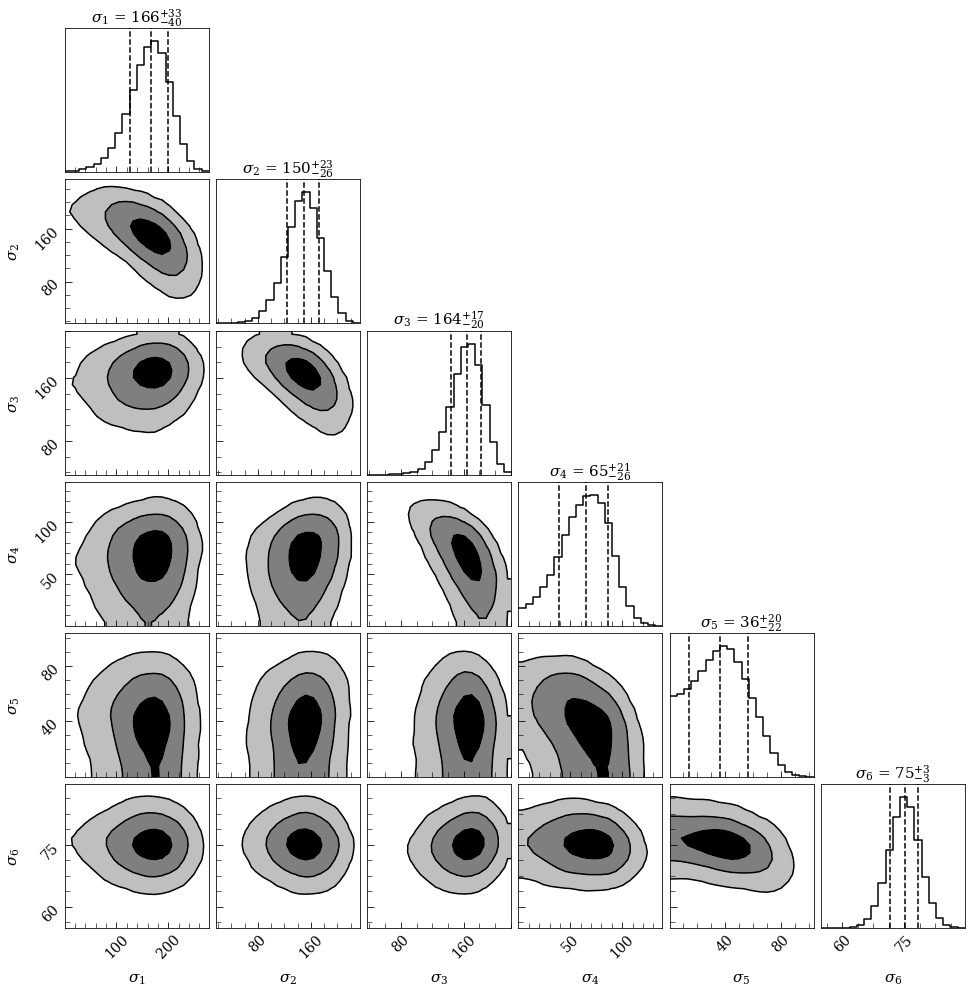}
     }
     \caption{Posterior distribution for the free parameters in the velocity dispersion map fitting. The dashed lines represent the 16th, 50th and 84th percentile. The units of the velocities are \kms.}
     \label{fig:corner_sigma}
 \end{figure*}

In figure \ref{fig:corner_flux} we present the corner plot distribution for the fitting of the flux map where we model the emitted flux as a sum of a 2D Sèrsic profile to take into account the extended emission from the galaxy and a 2D Gaussian to reproduce the emission from the marginally resolved and unresolved source of emission (BLR, outflow, bulge) which parameters are marked as ``AGN'' as it is the dominant contribution source.
In Figure \ref{fig:corner_velocity} and \ref{fig:corner_sigma} we present the posterior distribution for the kinematic fitting parameters presented in Section \ref{sec:kinematics}.


\subsection{Alternative kinematic fitting}

To understand if the velocity and velocity dispersion map fitting really require a nonparametric model to reproduce the observed galaxy kinematic, we also tried a more conservative approach adopting a parametric modeling of the velocity and velocity dispersion curves.

To model the velocity curve we assume that velocity contribution was only due to the exponential disk component that can be described as a function of radius:

\begin{equation}\label{velocityexpdisk}
    V_d^2(r) =  2 y^2 \frac{G M_D(R_0)}{r_D} \frac{\mathcal{I}_0(y)\mathcal{K}_0(y) - \mathcal{I}_1(y)\mathcal{K}_1(y)}{1- \exp(-R_0/r_D)(1 + R_0/r_D)}
,\end{equation}
where $y \equiv \frac{r}{2r_D}$, while $\mathcal{I}_0$, $\mathcal{I}_1$, $\mathcal{K}_0$ , and $\mathcal{K}_1$ are modified Bessel functions.
We fixed the value of $r_D$ as the value of the exponential disk scale radius found in the morphological fitting, and we left the disk mass $M_D$ computed within a radius $R_0 = 5$ kpc, which corresponds to $\sim$ 2-3 disk scale lengths: $M_d = M_d(R_0) = M_d(r = 5 {\rm kpc})$ free to vary.

The velocity dispersion curve was instead modeled as Gaussian:
\begin{equation}
    \sigma(r) = \sigma_0 \times \exp{-r^2/r_\sigma^2}
\end{equation}

where $\sigma_0$ is the velocity dispersion in the center of the galaxy and $r_\sigma$ is a measure of how quickly the velocity dispersion of the galaxy decreases with radius.

\begin{figure*}[ht!]
    \resizebox{\hsize}{!}
     { 
 	\includegraphics{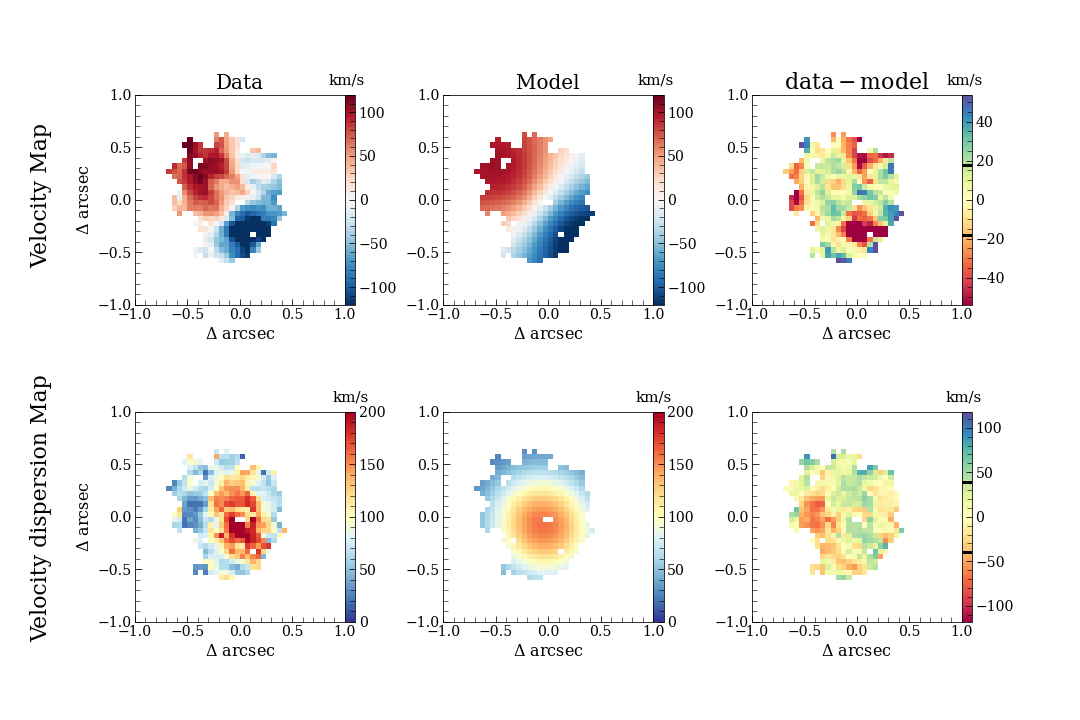}
     }
     \caption{
     Parametric best-fit results for the modeling of the \ha\ velocity and velocity dispersion maps. From left to right on the top the observed velocity map, the best-fit model map, and the residuals.
    From left to right on the bottom the observed velocity dispersion map, the best fitting model dispersion map, and the residuals.
    The colorbars of the residual range
between $-3\sigma$ and $+3\sigma$, and the black lines indicate $\pm 1\sigma$.
     }
     \label{fig:map_exponential}
 \end{figure*}

\begin{figure}[hb!]
    \resizebox{\hsize}{!}
     { 
 	\includegraphics{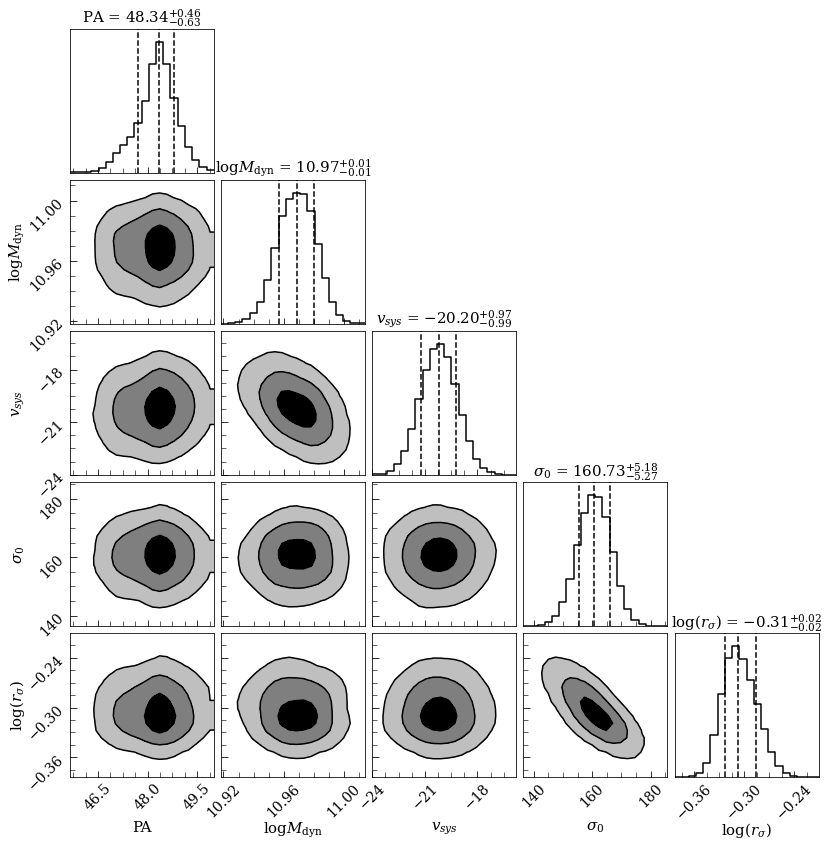}
     }
     \caption{
     Posterior distribution for the free parameters in the parametric fitting of the moment maps. The dashed lines represent the 16th, 50th, and 84th percentile. The PA is measured in degrees, the mass is measured in solar mass units, the systemic velocity and $\sigma_0$ in \kms, and $r_\sigma$ is measured in arcseconds.}
     \label{fig:corner_exponential}
 \end{figure}

We fixed the inclination at the value of 22 $\deg$, and we left the PA, disk mass, systemic velocity, $\sigma_0$, and $r_\sigma$ free to vary.  
The best-fit model and residuals are shown in Figure \ref{fig:map_exponential}, and the best-fit values and the posterior distribution are reported in Figure \ref{fig:corner_exponential}.

In Figure \ref{fig:map_exponential} the model velocity map is not able to reproduce the blue-shifted side. As the velocity in that side of the galaxy increases faster with radius than the red one, the exponential disk is not able to reproduce that fast growth, indicating the presence of an extra component in the central region of the galaxy to reproduce that fast rise in the velocity.
The BIC test for the velocity map extremely favors the nonparametric modeling ($\Delta BIC > 100$).

The Gaussian decreasing velocity dispersion model is also disfavored by the BIC test in comparison to the nonparametric modeling ($\Delta BIC \sim 20$), as it is not able to reproduce  the very high central velocity dispersions and the sharp decrease around 1.5 kpc.

\end{appendix}

\end{document}